\documentstyle[aaspp4]{article}

\def\ngc{NGC 6254}
\def\N{{\it N}}

\lefthead{Gnedin, Lee \& Ostriker}
\righthead{Tidal Shocks}

\begin{document}

\title{Effects of Tidal Shocks on the Evolution of Globular Clusters}

\author{Oleg Y. Gnedin}
\affil{Princeton University Observatory, Princeton, NJ~08544;\\
       ognedin@astro.princeton.edu}

\author{Hyung Mok Lee}
\affil{Department of Earth Sciences, Pusan University, Pusan 609-735 \\
       and Department of Astronomy, Seoul National University, Seoul 151-742,
       Korea;\\
       hmlee@astro.snu.ac.kr}

\author{Jeremiah P. Ostriker}
\affil{Princeton University Observatory, Princeton, NJ~08544;\\
       jpo@astro.princeton.edu}

\begin{abstract}
We present new Fokker-Planck models of the evolution of globular
clusters, including gravitational tidal shocks.  We extend our
calculations beyond the core collapse by adopting three-body binary
heating.  Effects of the shocks are included by adding the tidal shock
diffusion coefficients to the ordinary Fokker-Planck equation: the first
order heating term, $\langle\Delta E\rangle$, and the second order
energy dispersion term, $\langle\Delta E^2\rangle$.  As an example, we
investigate the evolution of models for the globular cluster NGC 6254.
Using the {\it Hipparcos} proper motions, we are now able to construct
orbits of this cluster in the Galaxy.  Tidal shocks accelerate
significantly both core collapse and the evaporation of the cluster and
shorten the destruction time from 24 Gyr to 18 Gyr.  We examine various
types of adiabatic corrections and find that they are critical for
accurate calculation of the evolution.  Without adiabatic corrections,
the destruction time of the cluster is twice as short.

We examine cluster evolution for a wide range of the concentration and
tidal shock parameters, and determine the region of the parameter space
where tidal shocks dominate the evolution.  We present fitting formulae
for the core collapse time and the destruction time, covering all
reasonable initial conditions.  In the limit of strong shocks, the
typical value of the core collapse time decreases from $10\, t_{rh}$ to
$3\, t_{rh}$ or less, while the destruction time is just twice that
number.  The effects of tidal shocks are rapidly self-limiting: as
clusters lose mass and become more compact, the importance of the shocks
diminishes.  This implies that tidal shocks were more important in the
past.
\end{abstract}

\keywords{stellar dynamics --- globular clusters: general ---
          globular clusters: individual (\ngc)}

\section{Introduction}

Evolution of star clusters is driven by a variety of dynamical
processes.  Two-body relaxation, stellar evolution, evaporation of
stars, and tidal truncation have all been shown to play a role.  In
addition to the internal effects, the external tidal shocks from the
Galaxy are very important.  \cite{GO:97} (1997) found that tidal shocks
contribute at least as much as two-body relaxation to the destruction of
the current Galactic sample of globular clusters.  \N-body simulations
of \cite{GO:99} (1999) provide a detailed account of the effect of
individual tidal shocks on the stellar energy distribution.  In this
paper we describe the implementation of tidal shocks in a Fokker-Planck
code and present new models of globular cluster evolution.

The study of the dynamical evolution of globular clusters has a long and
rich history.  \cite{S:87} (1987), \cite{MH:97}, and \cite{AZ:98}
provide comprehensive reviews of the evolution of isolated and
tidally-limited clusters.  Two-body relaxation causes loosely-bound
stars gain velocity higher than the escape velocity, which leads to
their evaporation from the cluster (\cite{A:38} 1938; \cite{S:40} 1940).
Even without evaporation, clusters would experience a catastrophic
collapse driven by the negative heat capacity and the conductive
transport of the thermal energy from the inner to the outer parts of the
cluster (\cite{A:62}; \cite{LW:68}; \cite{G:87}).  The run-away process
leads to a gravothermal catastrophe, or core collapse.  The collapse can
be reversed when a sufficient amount of heating is provided in the core
via formation of binaries.  The successive mergers of normal stars and
subsequent supernova explosions were also shown to be effective in
halting the collapse (Lee 1987).  After the reversal of core collapse,
the evolution of the cluster can be characterized by quasi-static
expansion with possible gravothermal oscillations.

The tidal field of the Galaxy affects cluster dynamics in two ways.  The
stellar distribution is effectively cut off at the point where the
ambient Galactic density exceeds that of stars in the cluster.
\cite{LO:87} found that clusters with short initial relaxation time
dissolve in $\sim N/200$ orbital periods around the Galaxy, where $N$ is
the initial number of stars in the cluster.  Tidal effects may also
cause cluster rotation as the stars on prograde orbits have higher
chances to be ejected than those on retrograde orbits (e.g.,
\cite{OL:92}).  In addition, the time-varying tidal forces cause
gravitational shocks when the cluster passes through the disk of the
Galaxy (\cite{OSC:72} 1972) or comes near the Galactic nucleus (e.g.,
\cite{S:87} 1987).  Tidal shocks increase random motion of stars and
reduce cluster binding energy.  Finally, \cite{KO:94} (1994) and
\cite{GO:97} (1997) emphasized that the tidal shock relaxation can be
very important in accelerating cluster evolution.  Recent studies
(\cite{MW:97} 1997a-c; \cite{OG:97}) show that these effects can lead to
substantial evolution of the globular cluster population.

All of the above processes are included in our Fokker-Planck code.  We
discuss the detailed implementation of tidal shocks and demonstrate
their effects on the evolution of clusters with a wide range of
parameters.  We consider both single-mass models and models including a
spectrum of stellar masses.  Mass segregation and rapid energy transfer
between the stars of different masses lead to an even faster core
collapse and destruction of globular clusters.  Although we do not
include the effects of stellar evolution, they are likely to be
important only at the early stages of cluster evolution.  \cite{CS:87}
and \cite{CW:90} showed that the mass loss from the short-lived massive
stars can disrupt weakly concentrated clusters (with $c \lesssim 0.6$).
We also ignore the effects of rotation and velocity anisotropy but
discuss the latter at the end of the paper.

We first review the theory of tidal shocks in \S 2.  The new features of
the Fokker-Planck code are described in \S 3.  We start with an
illustrative case that demonstrates the importance of tidal shocks and
adiabatic corrections in \S 4.  More systematic study for a wide range
of the parameter space is presented in \S 5.  Finally, we discuss the
implications of our results to the evolution of the Galactic star
clusters.


\section{Review of the Theory of Tidal Shocks}
\label{sec:theory}

The first and second order theory of the compressive (disk) shocks and
tidal (bulge) shocks has been developed by \cite{OSC:72} (1972);
\cite{S:87} (1987); and \cite{KO:94} (1994) and summarized in
\cite{GO:97} (1997).  Below we review the important results for the disk
shocking.  The theory of tidal shocking is discussed in detail in
\cite{GHO:99} (1999).  For an alternative linear theory calculation see
\cite{W:94}.

In the disk shocking, the phase-averaged first and second order energy
changes of stars with the initial energy per unit mass $E$ are
\begin{eqnarray}
\langle \Delta E \rangle_E & = & {2 \, g_m^2 \, r^2 \over 3 \, V^2} \; A_1(x),
  \label{eq:de_av} \\
\langle \Delta E^2 \rangle_E & = & {4 \, g_m^2 \, v^2 \, r^2 \,
    (1+\chi_{r,v}) \over 9 \, V^2} \; A_2(x),
  \label{eq:de2_av}
\end{eqnarray}
where $g_m$ is the maximum vertical gravitational acceleration produced
by the disk, $V$ is the vertical component of the cluster velocity with
respect to the disk, and $r$ and $v$ are the rms position and velocity
of stars of energy $E$.  Here $\chi_{r,v}$ is the position-velocity
correlation factor, which takes values from $-0.25$ to $-0.57$ (see
\cite{GO:99} 1999).

The first order energy change, $\langle \Delta E \rangle$, causes
reduction in the binding energy of the system and leads to evaporation
of the marginally bound stars.  The second order change, $\langle \Delta
E^2 \rangle$, causes a much larger energy dispersion which allows
additional stars to leave the cluster.  The two effects cooperate and
lead to faster dissolution of the cluster.

The adiabatic corrections, $A_1(x)$ and $A_2(x)$, account for
conservation of the adiabatic invariants of stars for which the orbital
period in the cluster is short compared to the effective duration of the
shock,
\begin{equation}
\tau \equiv {H \over V},
  \label{eq:tau}
\end{equation}
where $H$ is the characteristic scale-height of the disk.  The adiabatic
corrections can be approximately described as functions of the only
dimensionless parameter,
\begin{equation}
x \equiv \omega(r) \, \tau.
  \label{eq:xd}
\end{equation}
Here $\omega$ is the stellar orbital frequency of stars of energy $E$ at
the rms position $r$.

Several approximations have been used to study the adiabatic
corrections.  (1) The {\it impulse approximation} is valid for the stars
whose motion in the cluster is much slower than the shock, $x \ll 1$.
This condition applies in the outer regions of the cluster.  There both
corrections become unity:
\begin{equation}
A_1(x) = A_2(x) = 1, \;\;\;\;\; (x \ll 1).
\end{equation}
(2) The {\it harmonic approximation} is valid for the stars which
oscillate very quickly in a simple harmonic potential and which are not
in resonance with the perturbing force.  Building on results of
\cite{S:87} (1987), we get
\begin{equation}
A_{\rm 1S}(x) = e^{-2 x^2}, \;\;\;
A_{\rm 2S}(x) = {9 \over 4 + 5 e^{2 x^2}}, \;\;\;\;\;
   (x \gg 1; \mbox{~~no resonances}).
\label{eq:SpCorr}
\end{equation}
Here we have modified the second correction from its version in
\cite{GO:97} (1997) to make it equal unity for $x=0$.  The numeric
coefficients are dictated by the asymptotic form for large values of
$x$.  (3) The {\it linear theory} by \cite{W:94} includes the possible
resonances and predicts a less steep power-law function.  In the limit
of large $\tau$, the results can be approximated (\cite{GO:99} 1999)
as:\footnotemark

\footnotetext{This does {\it not} imply that the results of \cite{W:94}
  for fast shocks are incorrect.  The linear theory applies for all
  values of $\tau$, but the final expressions are complex and cannot be
  fitted simply.}

\begin{equation}
A_{\rm 1W}(x) = A_{\rm 2W}(x) = (1 + x^2)^{-3/2}.
\label{eq:WCorr}
\end{equation}
(4) \cite{GO:99} (1999) performed {\it \N-body simulations} of the tidal shock,
allowing for the self-consistent oscillations of the cluster potential as it
relaxes into a new virial equilibrium.  The simple fits to the results are
\begin{equation}
A_{\rm 1N}(x) = (1 + x^2)^{-5/2}, \;\;\; A_{\rm 2N}(x) = (1+x^2)^{-3}
\label{eq:NumCorr}
\end{equation}
for the ``fast shocks'' whose durations are comparable to or smaller than
the dynamical time at the half-mass radius: $\tau \lesssim t_{dyn}$.  For
the ``slow shocks'', the results agree with the predictions of the linear
theory (eq. [\ref{eq:WCorr}]).

In the case of bulge shocking, the Spitzer adiabatic corrections involve
a combination of Bessel functions, as described in \cite{GO:97} (1997).
The Weinberg and \N-body adiabatic corrections are taken to be the same
as for disk shocking.  Since the \N-body simulations of \cite{GO:99}
(1999) include only disk shocks, further study is needed to determine
accurately the adiabatic corrections for bulge shocking.

The first and second order energy changes are of comparable importance.
This becomes clear if we define the characteristic shock timescales:
\begin{mathletters}
\begin{equation}
t_{sh} \equiv {\left| E_h \right| \over dE_h/dt} = P_{\rm disk} \, 
    {\left| E_h \right| \over \langle \Delta E \rangle_h}
\end{equation}
\begin{equation}
t_{sh,2} \equiv {E_h^2 \over dE_h^2/dt} = P_{\rm disk} \, 
      {E_h^2 \over \langle \Delta E^2 \rangle_h}.
\end{equation}
\label{eq:definetimes}
\end{mathletters}
Here $P_{\rm disk}$ is the period of cluster's passage through the
Galactic disk.  The energy changes $\langle \Delta E \rangle_h$ and
$\langle \Delta E^2 \rangle_h$ are evaluated at the half-mass radius
$R_h$, and the characteristic energy $E_h$ is given by the Virial
Theorem: $\left| E_h \right| = v_{rms}^2/2 \approx 0.2GM/R_h$ (e.g.,
\cite{S:87} 1987), where $M$ is the mass of the cluster.  In the impulse
approximation we obtain
\begin{equation}
t_{sh} = {3 \over 4} \; P_{\rm disk} \; {V^2 \, \omega_h^2 \over g_m^2},
  \;\;\;\;
t_{sh,2} = {9 \over 16} \; P_{\rm disk} \; {V^2 \, \omega_h^2 \over g_m^2},
  \label{eq:tsh}
\end{equation}
where $\omega_h \equiv v_{rms}/R_h$ is the rms angular velocity of stars
at the half-mass radius.  The two timescales are simply related by
$t_{sh,2} = {3\over 4} \, t_{sh}$, so that both processes contribute
similarly to the destruction of the cluster.


\section{The Fokker-Planck Code}

We model the evolution of globular clusters using an orbit-averaged
isotropic Fokker-Planck code descended from Cohn (1979, 1980).  The code
has been modified by \cite{LO:87} and \cite{LFR:91} to include tidal
boundary and the three-body binary heating.  Stars beyond the tidal
boundary do not escape instantaneously but follow instead a continuous
distribution function $f(E)$, as described in \cite{LO:87}.  This
accounts for the balance of the internal and external forces at the
tidal radius.  Therefore, those stars only drift slowly away from the
cluster.  The tidal field is assumed to be spherically symmetric, which
is a weakness of the one-dimensional code (for a discussion see
\cite{LG:95} 1995).  We also assume that the cluster fills its Roche
lobe and require that the average density within the tidal radius remain
constant throughout the cluster evolution.

The heating of stars which reverses core collapse is provided by
three-body binaries.  They are included explicitly without following
their actual formation and evolution, according to the prescription by
\cite{C:85}.  \cite{O:85} argued that tidally-captured binaries are
probably more dynamically important for massive clusters.  Whether tidal
capture leads to a hard binary which contributes to stellar ejection
after a close encounter or whether it leads (more often) to a merger
which causes mass loss following stellar evolution, the outcome is the
same.  Lee (1987) showed that the merger of stars would give a nearly
identical dynamical effect on the cluster because the massive stars
formed in the merger would evolve off rapidly.  Both processes cause
indirect heating by ejecting mass.  The situation becomes more
complicated if we allow for the existence of primordial binaries and
massive degenerate stars.  The existence of massive remnant stars, such
as neutron stars, could cause three-body binaries to be more important
than tidal capture binaries (\cite{KLG:98}). However, many aspects of
the evolution after core collapse are independent of the actual energy
source (\cite{H:61} 1961; Goodman 1993).

We include the effects of tidal shocks by modifying the diffusion
coefficients in the Fokker-Planck equation.  We assume that the first
and second order energy changes are known as functions of the energy and
position using the results from \S \ref{sec:theory} and \cite{GHO:99}
(1999).  We now re-derive the Fokker-Planck equation in order to define
the diffusion coefficients corresponding to the shocks.

Let $\Psi(E,\Delta E)\ d\Delta E$ be the probability of scattering of a star
of energy $E$ by the amount $[\Delta E, \Delta E + d\Delta E]$.  \N-body
simulations (\cite{GO:99} 1999) show that the probability distribution is
nearly Gaussian:
\begin{equation}
\Psi(E,\Delta E)\ d\Delta E = {1 \over \sqrt{2\pi \langle\Delta E^2\rangle}} \;
  e^{-\frac{1}{2} {(\Delta E - \langle\Delta E\rangle)^2 \over
  \langle\Delta E^2\rangle}}\ d\Delta E,
  \label{eq:psi}
\end{equation}
where both $\langle\Delta E\rangle$ and $\langle\Delta E^2\rangle$ are
functions of energy $E$.  Let $N(E) dE$ be the number of stars in the
energy range $[E, E+dE]$.  The time evolution of $N(E)$, in the absence
of two-body relaxation and binary heating, can be described by
the following equation
\begin{equation}
N(E, t+\Delta t) = \int \; N(E-\Delta E, t) \, \Psi(E-\Delta E, \Delta E)\
   d\Delta E,
\end{equation}
where the energy changes by the amount $\Delta E$ in the time interval
$\Delta t$.  Expanding this equation in a series over $\Delta E$ and
$\Delta t$, we obtain the usual form of the Fokker-Planck equation
\begin{equation}
{\partial N(E)\over \partial t} = -{\partial\over \partial E}
  \left\{ N(E) \left<{\cal D}_t(\Delta E) \right>_V\right\} + 
  {1\over 2} {\partial^2\over \partial E^2}
  \left\{ N(E)\left<{\cal D}_t (\Delta E^2)\right>_V\right\},
\end{equation}
where $\left<{\cal D}_t(\Delta E)\right>_V$ and
$\left<{\cal D}_t(\Delta E^2)\right>_V$ are the orbit-averaged diffusion
coefficients of the first and second order.  They are defined by
\begin{equation}
\left<{\cal D}_t (\Delta E)\right>_V \equiv {\int_0^{r_{max}} \left<{\cal D}_t 
  (\Delta  E)\right> vr^2 dr \over \int_0^{r_{max}} vr^2 dr},
\end{equation}
and
\begin{equation}
\left<{\cal D}_t(\Delta E)^2\right>_V \equiv {\int_0^{r_{max}}
  \left<{\cal D}_t(\Delta E^2)\right> vr^2 dr \over \int_0^{r_{max}} vr^2 dr},
\end{equation}
where $\left<{\cal D}_t(\Delta E)\right>$ and $\left<{\cal D}_t(\Delta
E^2)\right>$ are the average rates of change of the energy and its
dispersion, respectively, per unit volume at a given location $r$, and
$r_{max}$ is the maximum radius allowed for a star of energy $E$.  Given
equations (\ref{eq:de_av}) and (\ref{eq:de2_av}) for the energy changes
and equation (\ref{eq:psi}) for their distribution function, we define
\begin{eqnarray}
\left<{\cal D}_t(\Delta E)\right>_{sh} & \equiv &
  \int {\Delta E \over \Delta t} \, \Psi(E,\Delta E)\ d\Delta E
  \; = \; {\left<\Delta E\right> \over \Delta t}, \\
\left<{\cal D}_t(\Delta E^2)\right>_{sh} & \equiv &
  \int {(\Delta E)^2 \over \Delta t} \, \Psi(E,\Delta E)\ d\Delta E
  \; = \; {\left<\Delta E^2\right> + \left<\Delta E\right>^2 \over \Delta t}.
\end{eqnarray}

The bulge and disk tidal shocks are applied at the time step when the
cluster is at its perigalacticon or crosses the Galactic disk,
respectively.  The diffusion coefficients are normalized to the current
time step, $\Delta t$, so as to produce the right amount of heating per
event.  The shocks are repeated every orbital period of the cluster.

\begin{figure} \plotone{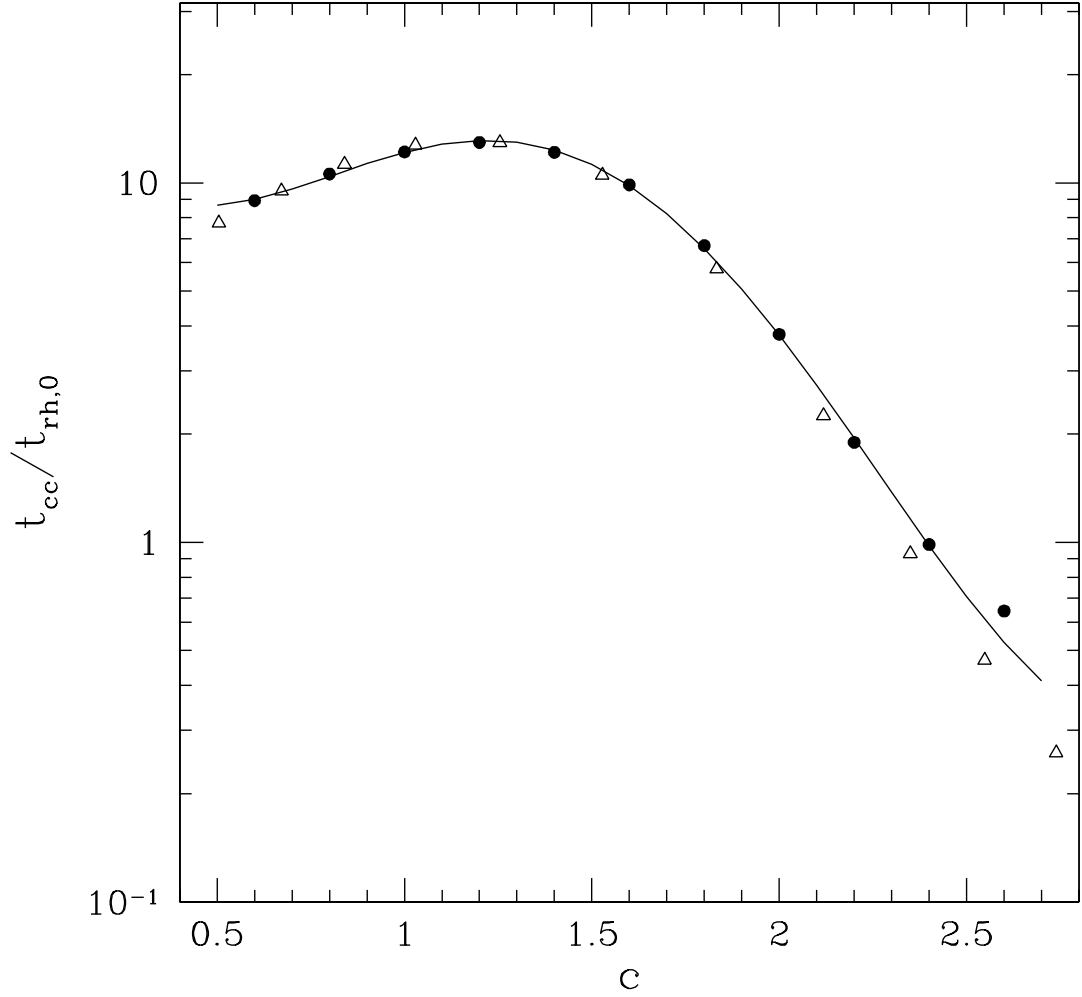}
\figcaption{The core collapse time as a function of the initial
  concentration for the models without tidal shocks.  Filled dots are
  our results, triangles are from \protect\cite{Q:96}.  The agreement is
  very good, except for the very high concentration clusters.  The solid
  line is our fit, eq. (\protect\ref{eq:tcc_rel}), with the last point
  excluded.  \label{fig:tcc_rel}}
\end{figure}

The time step was chosen such that both the cluster mass and the central
density change by no more than 1\% between time steps:
\begin{equation}
\Delta t = 0.01\times \min[(d\ln{\rho}/dt)^{-1}, (d\ln{M}/dt)^{-1}].
  \label{eq:dt}
\end{equation}
\cite{LO:93} suggested that such a time step should be sufficient for an
accurate calculation of the cluster evolution, both before and after core
collapse.  We compare our results without tidal shocking with those of
\cite{Q:96}, who used the same time step for the potential re-computation but
broke it down into 32 sub-steps for updating the distribution function (the
``Fokker-Planck steps'').  We found a measurable difference in the core
collapse times only for very concentrated clusters, with the initial
concentration $c \gtrsim 2.5$.  For these clusters an even smaller time step
may be required, although our results never differ by more than 30\%
(cf. Figure \ref{fig:tcc_rel}).

As described below, in order to incorporate correctly the effects of
tidal shocks into the code, the shock diffusion coefficients should be
applied separately from the relaxation terms, and the potential should
be recomputed accordingly thereafter (an alternative approach has been
suggested by \cite{JHW:98}).  It seems awkward, though still possible,
to include a set of Fokker-Planck steps for every potential time step in
this scheme, as \cite{Q:96} did for the isolated clusters.  We chose not
to do so and therefore we had the potential updated every time the
distribution function was changed.  Since our time step
(eq. [\ref{eq:dt}]) gives very good accuracy for most of the computed
cluster models, we did not change it specifically for high concentration
clusters.  The resulting error should in any case be small.

The alternative implementation of tidal shocks in a Fokker-Planck code
has been done by \cite{MW:97} (1997a-c).  Those authors employ the
linear theory formalism of \cite{W:94} and calculate the change of the
distribution function directly from the Boltzmann equation.  That
approach is more accurate but is more cumbersome.  Since the linear
theory expressions are supported by the restricted \N-body simulations
(Weinberg 1994), that method and our derived adiabatic corrections
should give the same result.  The work of \cite{MW:97} has focused more
on the destruction time of globular clusters under various assumptions
and on predicting the initial cluster distribution.  In this paper, in
addition to obtaining fitting formulae for the destruction time we
present also the detailed evolution of individual clusters and compare
it with previous work on isolated clusters.

\subsection{Comparison with \N-body simulations}

\cite{GO:99} (1999) report self-consistent \N-body simulations of a
single disk shock of various strengths and durations.  The cluster is
modeled as a low-concentration King-model ($c=0.84$) with a realistic
number of particles, $N=10^6$.  The characteristic disk shocking
strength is low and most of the stars remain bound to the cluster.
However, the energy distribution changes in response to the
perturbation.  The resulting distribution after the virialization phase
is accurately fitted by equations (\ref{eq:de_av}), (\ref{eq:de2_av}),
and (\ref{eq:NumCorr}).  Although the result depends on the initial
structure of the cluster, the above fits still apply for the more
concentrated cluster with $c=1.5$ (see \cite{GO:99} 1999 for more
details).

The \N-body modeling of tidal shocks is intrinsically more accurate than
the Fokker-Planck formulation.  The former takes into account the
non-linear virialization following the shock and the anisotropic effects
induced by disk shocking.  A direct comparison of the resulting energy
changes between the two codes is difficult because of the algorithmic
differences.  Therefore, we have taken the simple approach to reproduce
the \N-body results in our F-P code as closely as possible.

The general procedure for advancing the distribution function (DF) in
time consists of two parts (Cohn 1979).  The first part introduces the
first and second order changes of the DF as a solution of the
Fokker-Planck equation.  The second part updates the cluster potential
by solving Poisson's equation with the density following from the new
DF.  During the second step, the DF is kept constant as a function of
the stars' adiabatic invariants and therefore it changes as a function
of energy.  This procedure is no longer correct when treating tidal
shocks which last of the order of the dynamical time of the cluster.
This timescale is short enough that adiabatic invariants may themselves
change.

\begin{figure} \plotone{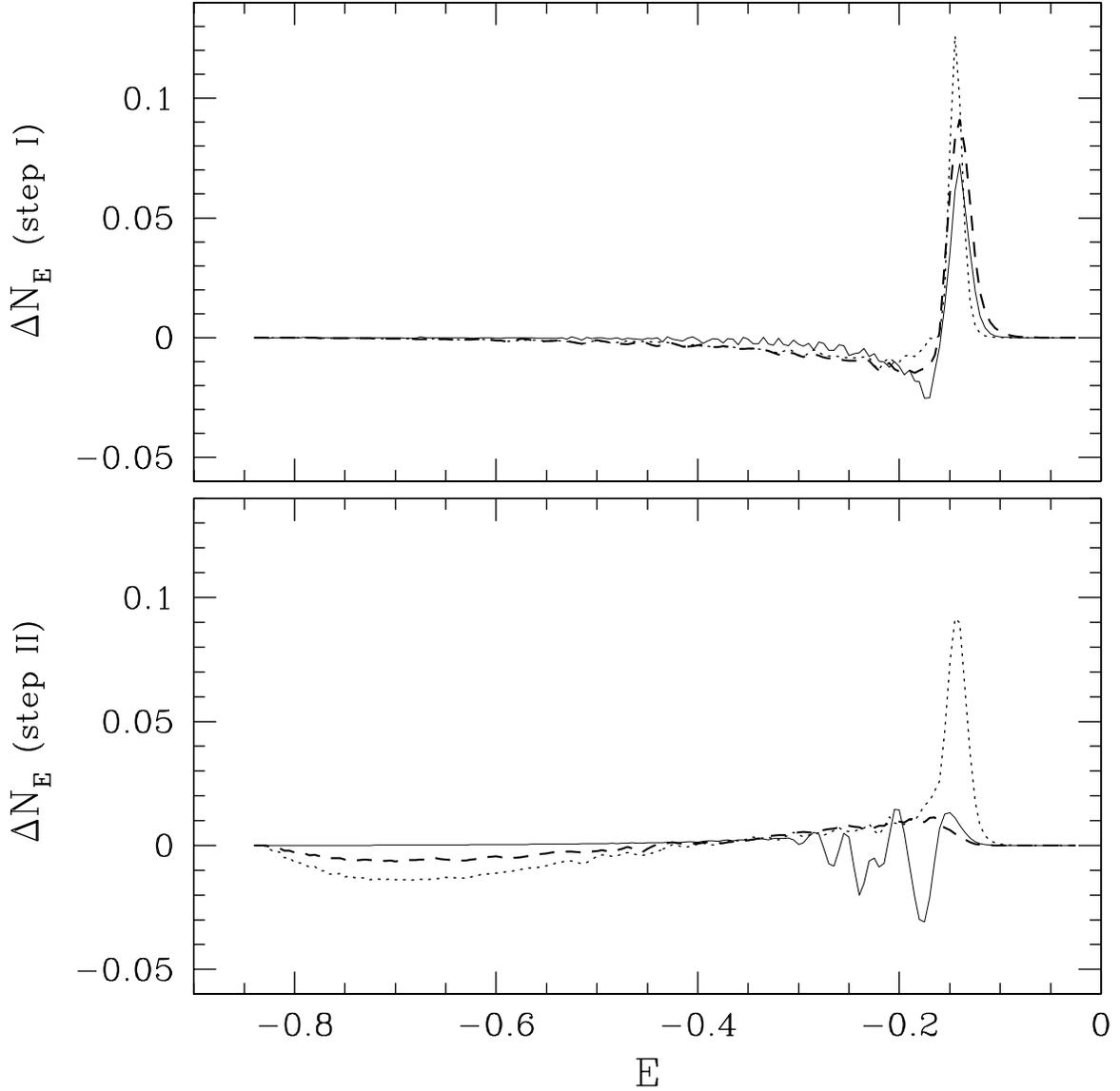}
\figcaption{Comparison of the change of the energy density, $\Delta
  N(E)$, for the current Fokker-Planck models ({\it dashes}) and the
  \N-body calculations ({\it solid lines}).  Energy on the horizontal
  axis is normalized to the cluster binding energy.  The test cluster is
  initially a King model with the concentration $c=0.84$.  A single weak
  shock is applied, similar to the ones studied by \cite{GO:99} (1999).
  {\it Top panel:} the energy change due to the shock; {\it Bottom
  panel:} the energy change due to the potential readjustment.  Dots
  show previous implementation of tidal shocks in the F-P code, where
  the distribution function was kept fixed as a function of adiabatic
  invariants.  The current procedure agrees with the \N-body results
  more accurately, especially in the center and in the outer parts of
  the cluster.  \label{fig:ne}}
\end{figure}

We find that in order to reproduce the \N-body results, we need to
modify the above procedure for tidal shocks.  During the time step when
the cluster passes through the Galactic disk or close to the Galactic
center, we first apply the shock diffusion coefficients and update the
DF.  Then we fix the DF as the function of energy and obtain the new
potential by solving Poisson's equation.  Specifically, since the
potential is changed and the energy grid is renormalized, we require
that in each energy bin the new energy distribution, $N_{new}$, scale
with the new energy grid, $E_{new} = E + \Delta E$, as
\begin{equation}
N_{new}(E_{new}) \, dE_{new} = N(E) \, dE,
\end{equation}
where $N(E)$ is the energy distribution before the potential
recalculation.  We have checked that the above procedure reproduces well
the \N-body results for a range of shock amplitudes.  Figure
\ref{fig:ne} shows the change in the energy distribution after a single
shock in a test cluster.  The agreement between the Fokker-Planck and
\N-body results is reasonably good.  Also, Figure \ref{fig:ne} shows
that the previous procedure for updating the DF for the fixed adiabatic
invariants departs much more strongly from the \N-body results than does
our new prescription.

Another effect demonstrated by the \N-body simulations is a slight
decrease in the shock-induced energy dispersion subsequent to and caused
by the potential fluctuations following the shock.  This paradoxical
``refrigeration effect'' persistently survived all our tests and seems
to be real (see \cite{GO:99} 1999).  However, the magnitude of the
dispersion decrease is relatively small, which led us to ignore it in
the present calculation.  Once the nature of this effect is understood
better, it should also be included in the models of globular clusters.


\section{Typical Evolution of a Single Cluster}

We consider now the detailed models of the globular cluster evolution.
For the most part, we study single-mass models and then briefly
investigate the effect of a spectrum of stellar masses.

\begin{figure} \plotone{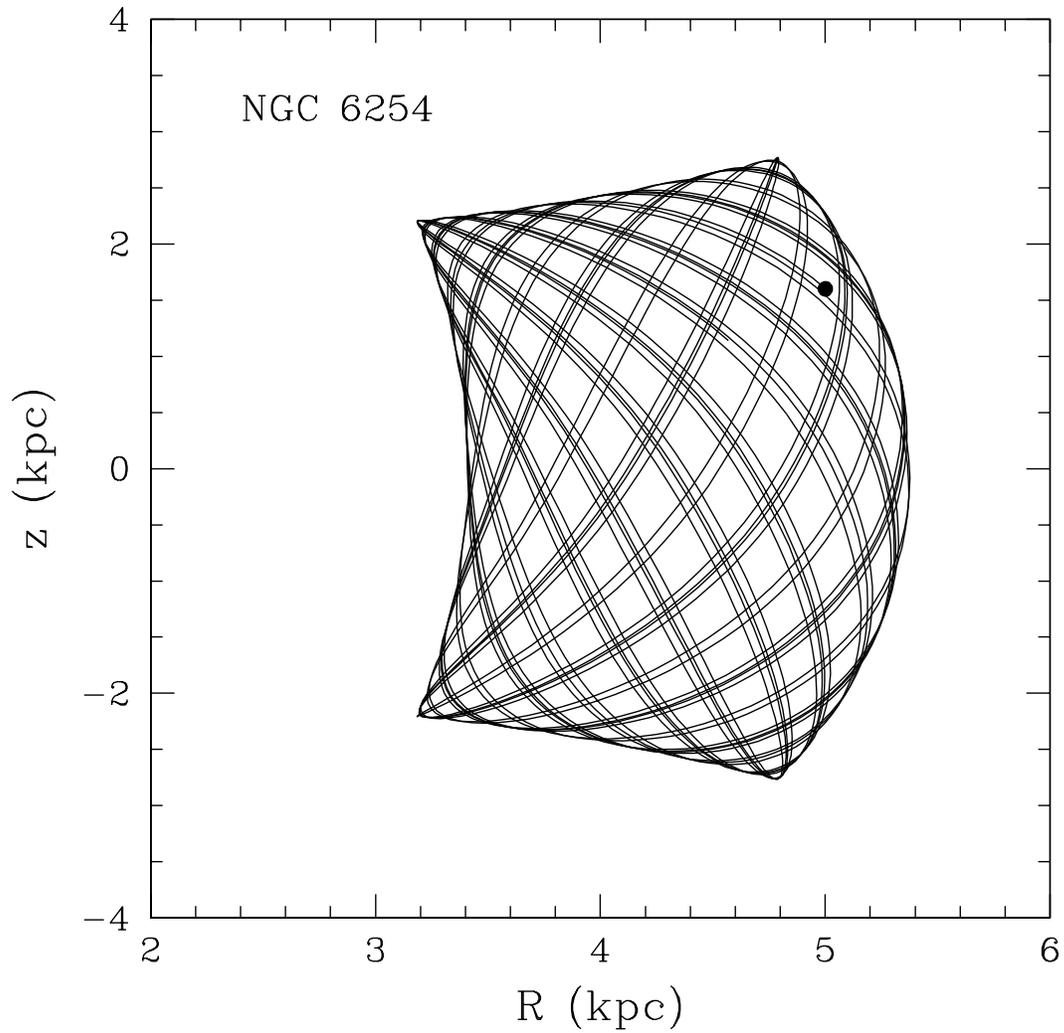}
\figcaption{Orbits of \ngc\ in the analytic potential of the Galaxy from
  \cite{AS:91}, for a period of $3\times 10^9$ years.  The dot marks the
  current position of the cluster.  \label{fig:orbits}}
\end{figure}

With the new data on proper motions from {\it Hipparcos}, it is now
possible to reconstruct full three-dimensional velocities for some of
the Galactic globular clusters (\cite{Oetal:97}) and to build realistic
models of their evolution.  We use the inferred orbital parameters for
NGC 6254 and calculate its orbits in an analytic potential of the Galaxy
from \cite{AS:91}.  The orbits are of the rosette type (Figure
\ref{fig:orbits}) with a small eccentricity, $e=0.21$.  The cluster
moves around the Galaxy with a period of about $1.4\times 10^8$ yr,
during which it passes next to the Galactic center every $9.5\times
10^7$ yr and crosses the Galactic disk every $5.3\times 10^7$ yr.  The
period ratio differs from the expected factor of two due to the slow
precession of the orbit.

From the computed orbits of \ngc\ we infer that the amplitude of tidal
shocks varies approximately as a Gaussian function of time.  This
satisfies the conditions under which the \N-body adiabatic corrections
were calculated (see \cite{GO:99} 1999).

\subsection{Single Mass Models}

The observed structure of \ngc\ is well characterized by a King model
(\cite{K:66}) with the concentration $c=1.4$, or the structural
parameter $W_0=6.55$.  The mass of the cluster is $2.25 \times 10^5 \,
M_{\sun}$, assuming a constant mass-to-light ratio $M/L_V = 3$ in solar
units.  First, we consider in detail a model composed of stars of the
same mass, $m_{*} = 0.7\, M_{\sun}$.  The important parameters of the
model are summarized in Table \ref{tab:param}.

We implement both disk and bulge shocks in our integration.  The models
differ in the way the shocks are included: no tidal shocks, steady tidal
field and relaxation only ({\it case 0}); tidal shocks with the first
order $\langle\Delta E\rangle$ term only ({\it case 1}); tidal shocks
with the second order $\langle\Delta E^2\rangle$ term only ({\it case
2}); and finally tidal shocks with both $\langle\Delta E\rangle$ and
$\langle\Delta E^2\rangle$ terms ({\it case 3}).

\begin{figure} \plotone{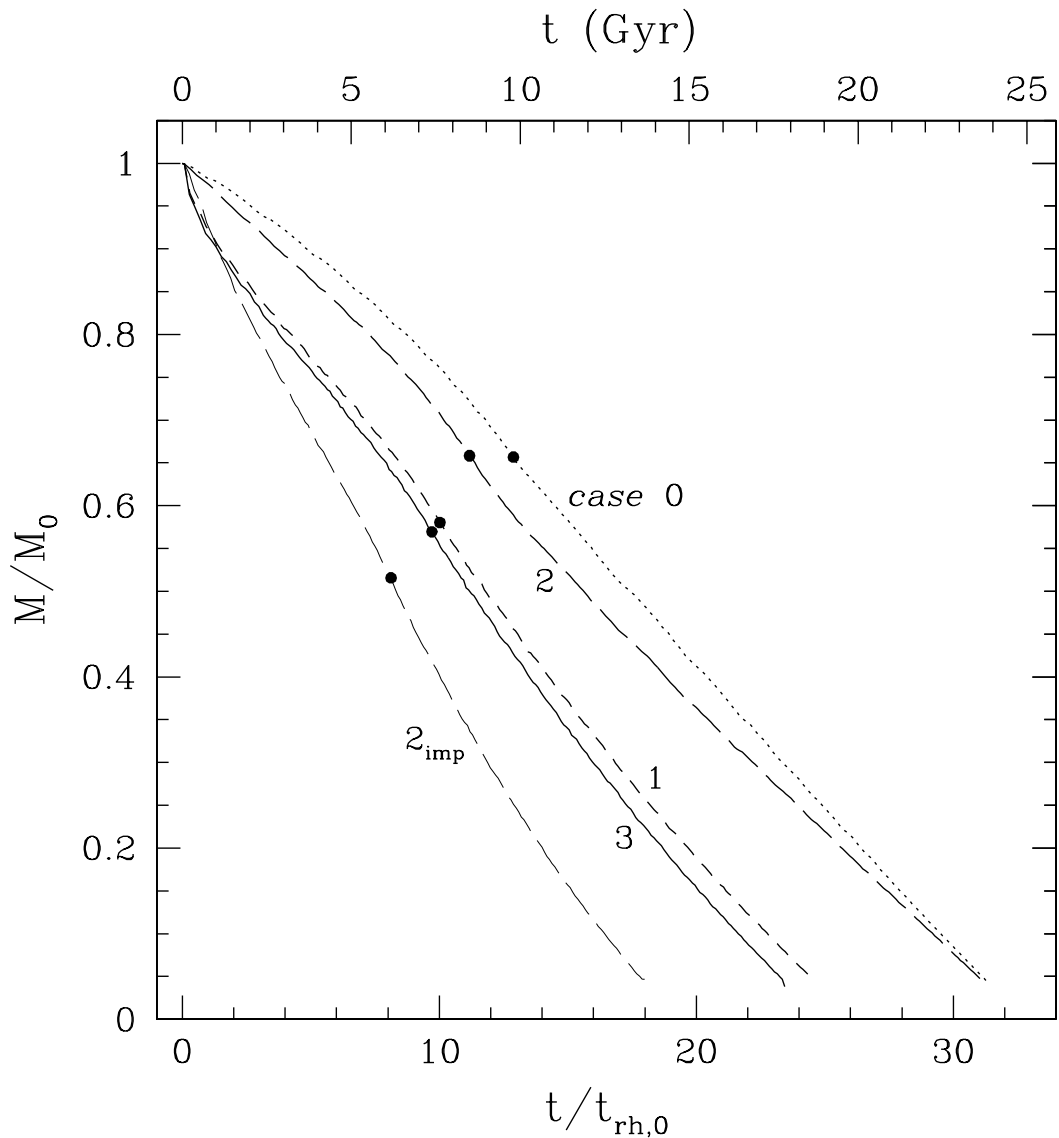}
\figcaption{The mass loss for the single-mass models of globular cluster
  \ngc.  Time is expressed in units of the initial half-mass relaxation
  time, as well as in billions of years.  Dots are for the model with
  two-body relaxation only ({\it case 0}), dashes are for the model that
  includes the effects of the energy shift due to tidal shocks ({\it
  case 1}), long dashes are for the model that includes the second order
  energy dispersion (without the heating term; {\it case 2}), and the
  solid line is for the final model including proper treatment of all of
  the shock effects ({\it case 3}).  The adiabatic corrections for
  ``slow shocks'' are used (eq. [\protect\ref{eq:WCorr}]).  For
  illustration, thin dashes show the effect of the second order term
  (marked ``{\it 2}$_{\rm imp}$'') if it had not had adiabatic
  corrections.  Filled circles indicate the time of maximum core
  collapse.  \label{fig:m}}
\end{figure}

An important observable of the cluster evolution is the mass-loss due to
tidal shocks and due to the direct escape of high-velocity stars.
Figure \ref{fig:m} shows the run of the cluster mass with time.  As is
well-known, the mass of a tidally limited cluster goes to zero in an
approximately linear fashion.  A convenient choice is to express time
$t$ in the units of the initial half-mass relaxation time $t_{rh,0}$
(eq. [\ref{eq:trh}]).  Two-body relaxation leads to destruction of the
cluster in about $32\, t_{rh,0}$.  Note that the numerical problems
cause the cluster to lose its stability and dissolve into the sea of
background stars before its mass reaches exactly zero.  Usually the code
fails to re-calculate the cluster potential when the mass falls to about
1\% -- 4\% of the initial mass.  However, we can extrapolate through the
last few time steps in order to estimate the hypothetical time when the
cluster mass vanishes.  The disruption time obtained this way is an
overestimate of the real disruption time, but all other estimates would
suffer from a personal choice.  Thus, in the rest of the paper we use
the extrapolation of our calculations as the destruction time $t_d$.

The destruction time is significantly reduced when we include
gravitational shocks.  The $\langle\Delta E\rangle$ term alone changes
$t_d$ from $32\, t_{rh,0}$ to $26\, t_{rh,0}$ and the shock-induced
relaxation brings it to about $24\, t_{rh,0}$.  To access the importance
of the two shock terms separately, we turn off the energy shift,
$\langle\Delta E\rangle$, for the {\it case 2} run.  The second order
term leads to an enhanced mass loss relative to the relaxation case for
most of the evolution.  This effect is not as strong as the first order
effect due to the limiting adiabatic corrections, which we investigate
in more detail later\footnotemark.  For comparison, we also show the
mass loss that would have been due to the shock-induced relaxation if
there were no adiabatic corrections (the impulse approximation).  The
latter effect would reduce the destruction time of the cluster by about
a half.

\footnotetext{The effect of the second order term is reduced relative to the
  results of \cite{GO:97} (1997) by inclusion of the correlation factor
  $(1+\chi_{r,v})$ (eq. [\ref{eq:de2_av}]).  At each time step, we calculate
  the value of $\chi_{r,v}(E;c)$ using the fitting formula from \cite{GO:99}
  (1999).}

An important feature of tidal shocks is the enhanced mass loss at the
early stages of the evolution, prior to core collapse.  In the pure
relaxation case, the mass loss is slow until core collapse and linear in
time afterwards.  The first theoretical calculations of the probability
of stars' escape from the cluster through two-body relaxation
(\cite{A:38} 1938; \cite{S:40} 1940) predicted the following
dimensionless rate
\begin{equation}
\xi_e \equiv - {t_{rh}(t) \over M(t)} \, {dM \over dt} = 0.0074,
\end{equation}
where $t_{rh}$ is the half-mass relaxation time (\cite{SH:71}):
\begin{equation}
t_{rh} = 0.138 \, {M^{1/2} \, R_h^{3/2} \over G^{1/2} \, m_* \, \ln(\Lambda)}.
  \label{eq:trh}
\end{equation}
Here $M$ is the current cluster mass, $R_h$ is the half-mass radius,
$m_*$ is the average stellar mass, and $\ln(\Lambda) = \ln(0.4N)$ is the
Coulomb logarithm, $N$ being the number of stars in the cluster.  Later
calculations of the tidally-truncated cluster by \cite{H:61} (1961) gave
a larger value, $\xi_e \approx 0.045$.

\cite{SC:73} used Monte-Carlo simulations for several models of globular
clusters and found $\xi_e=0.05$ for $R_t/R_h=3.1$, and $\xi_e=0.015$ for
$R_t/R_h=9.3$, where $R_t$ is the tidal radius of the cluster.  Based
upon those simulations, \cite{AHO:88} proposed the following fitting
formula for the time to disruption:
\begin{equation}
t_d = \left( 0.15 \, {R_h \over R_t} \right)^{-1} \, t_{rh,0},
\label{eq:tevAHO}
\end{equation}
where $t_{rh,0}$ is the initial relaxation time.  For \ngc, the
disruption time predicted by AHO would be about $52\, t_{rh,0}$, whereas
we see that the cluster dissolves in $32\, t_{rh,0}$ even without tidal
shocks.  In general, we find that equation (\ref{eq:tevAHO})
overestimates the actual disruption time by a factor of several.

\begin{figure} \plotone{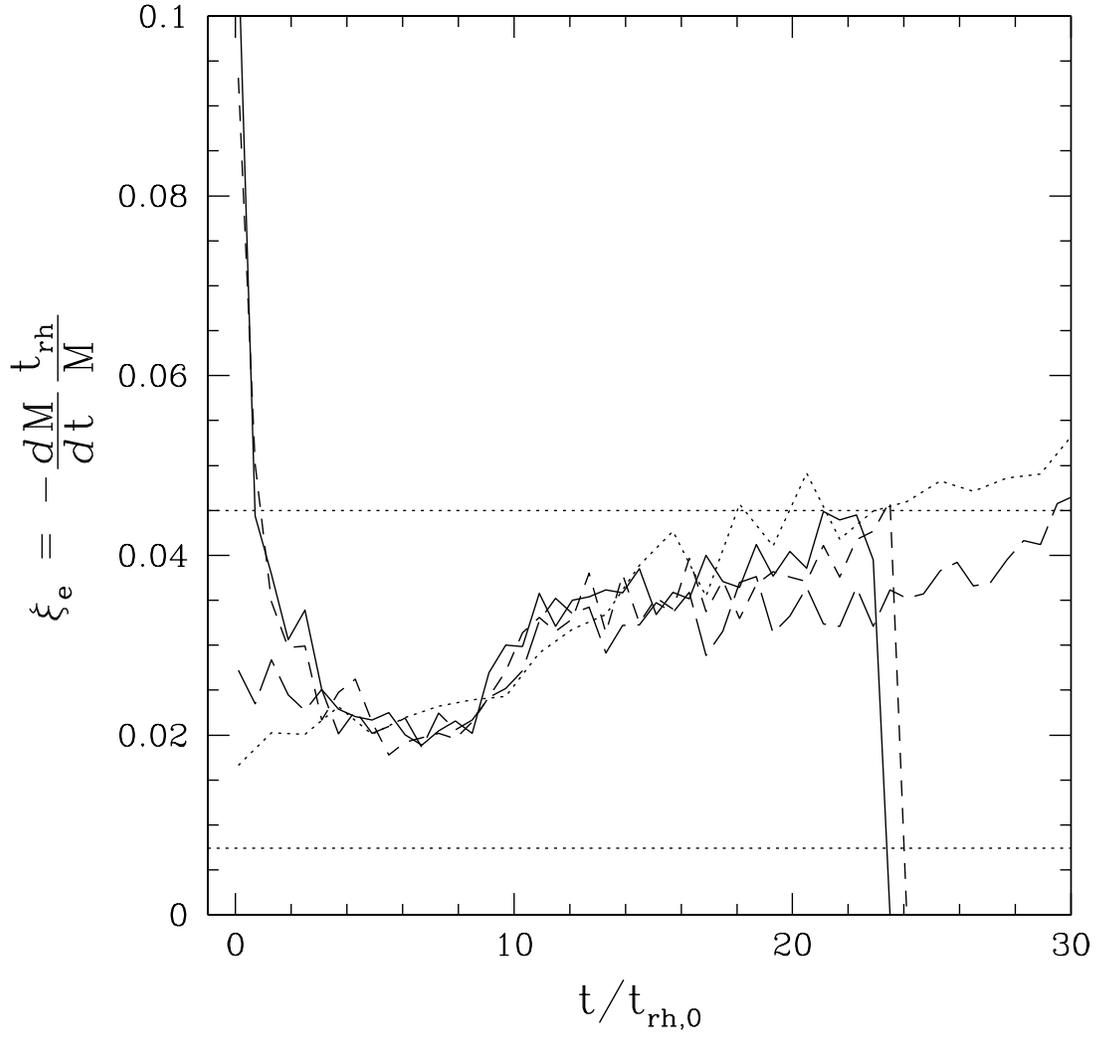}
\figcaption{Escape probability in units of the half-mass relaxation time
  for \ngc.  Line notation is the same as in Fig. \protect\ref{fig:m}.
  The lower horizontal line shows the analytical estimate by
  \protect\cite{A:38} (1938), $\xi_e = 0.0074$, and the upper
  self-similar calculations by \protect\cite{H:61} (1961), $\xi_e =
  0.045$.  \label{fig:dm}}
\end{figure}

We compare the previous results for the escape probability $\xi_e$ with
our Fokker-Planck calculations in Figure \ref{fig:dm}.  In the
relaxation model ({\it case 0}), the value of $\xi_e$ rises almost
monotonically through core collapse and until destruction of the
cluster, reaching and exceeding H\'{e}non's estimate only in the late
stages of the evolution.  On the contrary, in the shock-dominated models
the escape probability is very high in the beginning when tidal shocks
efficiently remove stars from the outer parts of the cluster.  The mass
loss then slows down and conforms to the relaxation rate soon before
core collapse.  Then again, $\xi_e$ rises with time to reach H\'{e}non's
self-similar value.  This later increase caused primarily by the
declining mass of the cluster, since the mass loss rate is almost
constant in the post core-collapse cluster.

\begin{figure} \plotone{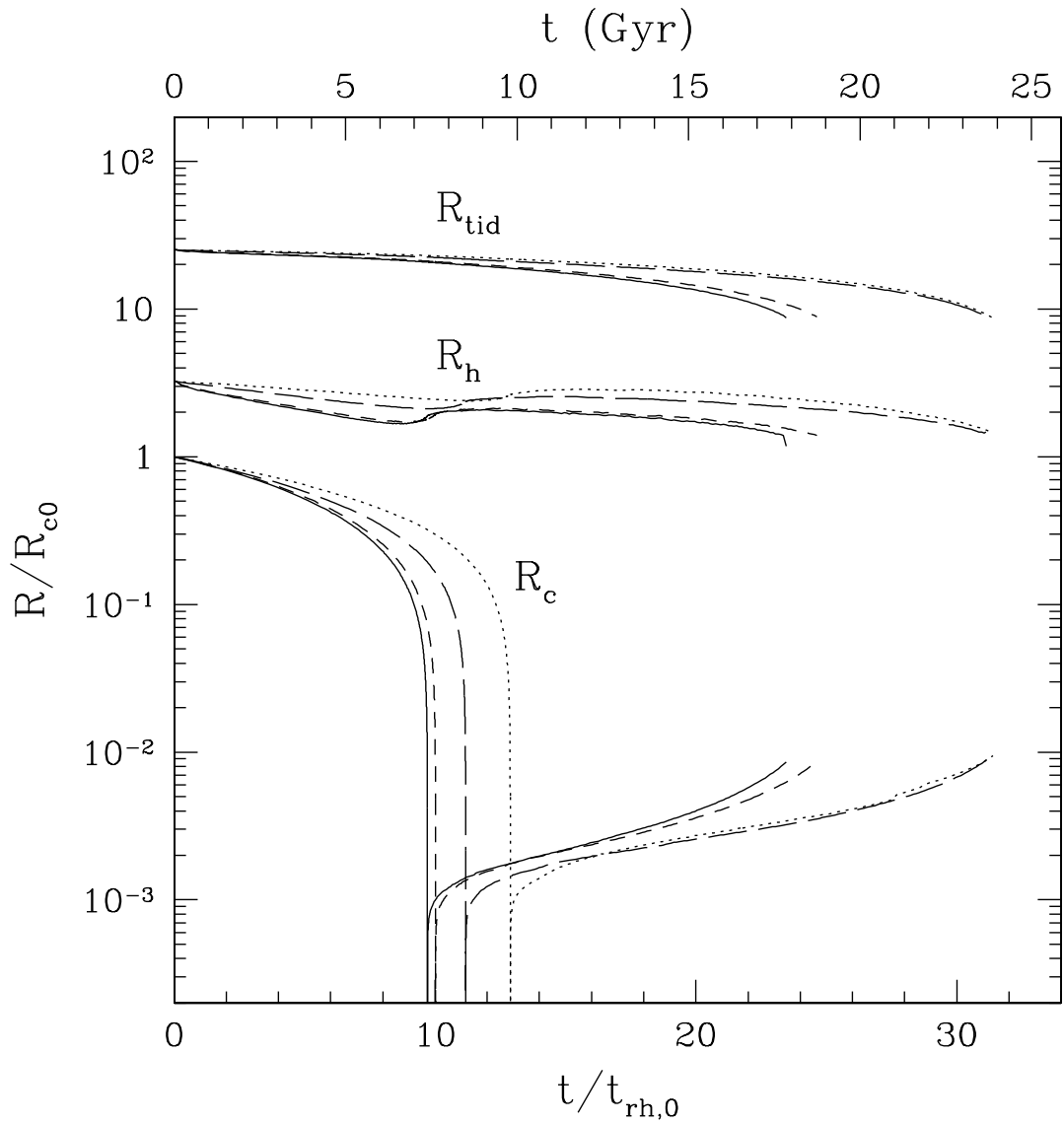}
\figcaption{Evolution of the tidal, half-mass, and core radii of \ngc.
  Line notation is the same as in Fig. \protect\ref{fig:m}.
  \label{fig:r}}
\end{figure}

The cluster structure can be described in terms of the characteristic
radii, such as the {\it core radius}, reflecting compactness of the
cluster; the {\it tidal radius}, confining all stars bound to the
cluster; and the {\it half-mass radius}, relating to the global
properties of the cluster.  Figure \ref{fig:r} shows the evolution of
these parameters normalized to the initial core radius.  The core radii
for all models drop to extremely small values at the point of core
collapse.  The core-collapse time $t_{cc}$ is about $13\, t_{rh,0}$ for
the relaxation case, in agreement with \cite{LO:87} and \cite{Q:96}.
Tidal shocks speed up core collapse significantly: $t_{cc} = 10\,
t_{rh,0}$ in {\it case 1} and it is still smaller when we include the
shock-induced relaxation term.  The half-mass radius changes only by a
factor of two over the whole cluster lifetime, which indicates that core
collapse affects only a small fraction of stars.  The tidal radius
decreases slowly, reflecting the gradual mass loss from the cluster.

\begin{figure} \plotone{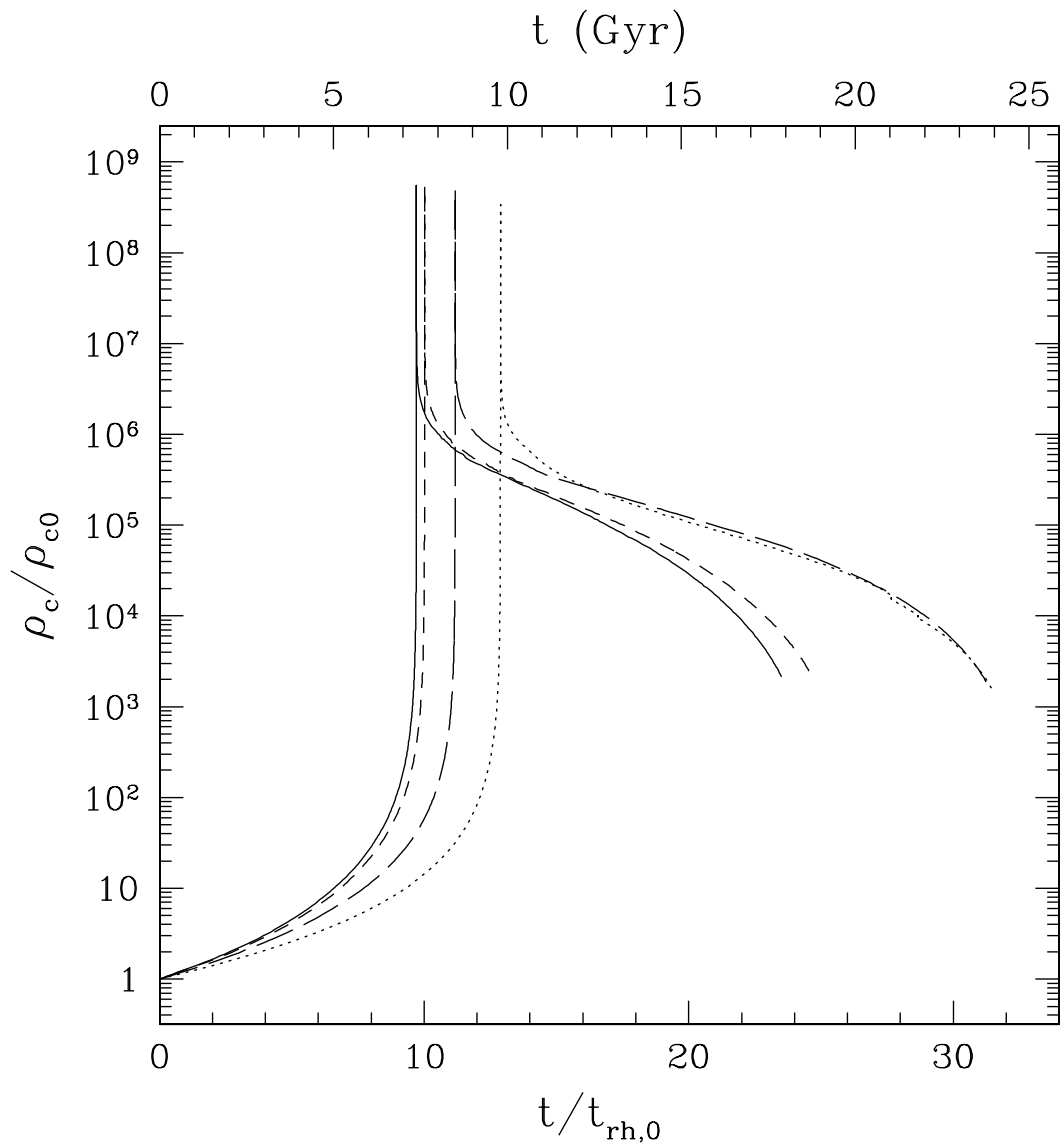}
\figcaption{Evolution of the central density of \ngc.
  Line notation is the same as in Fig. \protect\ref{fig:m}.
  \label{fig:rho}}
\end{figure}

Figure \ref{fig:rho} illustrates core collapse more clearly.  The
central density of the cluster, $\rho_c$, rises by nine orders of
magnitude before the collapse is halted by a production of three-body
binaries in the very dense core of the cluster.  The energy released
from the binaries eventually reverses the collapse.  The cluster
consequently expands, though the central density still remains very high
(about four orders of magnitude higher than initially).  The actual
value of $\rho_c$ at the core bounce and post-collapse phase depends on
the specific heating mechanism.  A stronger heating rate gives a lower
central density during the post-collapse phase.

The cluster concentration, $c \equiv \log_{10}(R_t/R_c)$, is closely
related to the central density.  Since the tidal radius $R_t$ decreases
slowly with time as $M^{1/3}$, the concentration varies roughly as
$\log{\rho_c}$.

\begin{figure} \plotone{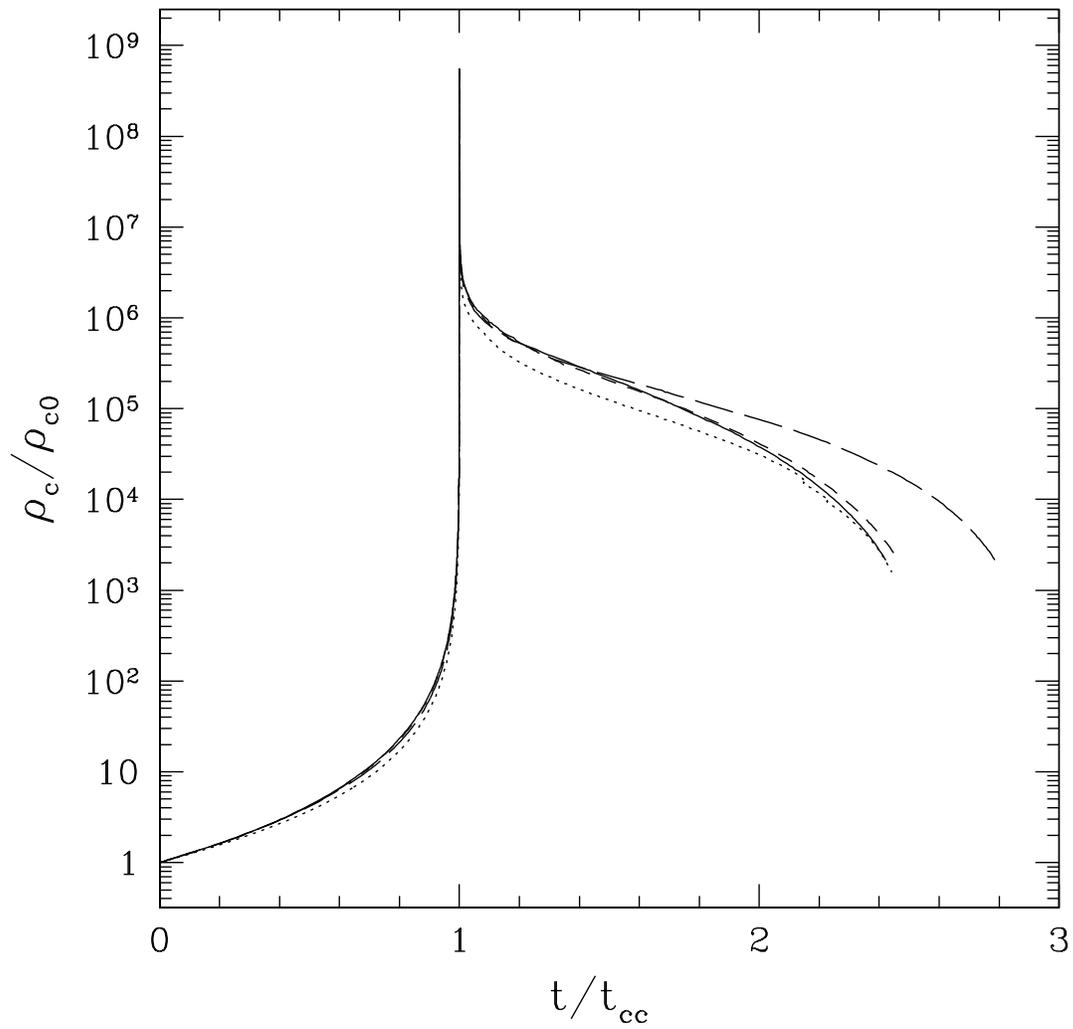}
\figcaption{Central density of \ngc\ vs. time scaled to the individual
  core collapse times.  Evolution in these scaled units is similar for
  all models.  Line notation is the same as in Fig. \protect\ref{fig:m}.
  \label{fig:rho_cc}}
\end{figure}

The core collapse time provides a very useful yardstick for cluster
evolution.  Even in the presence of tidal shocks, the evolution of the
models is similar when the units of time are scaled to the core collapse
time.  Figure \ref{fig:rho_cc} illustrates this point.  The central
density in models {\it 0--3} varies almost identically with $t/t_{cc}$.
Therefore, we can use this scaling to parameterize cluster evolution.
The destruction time is an almost constant multiple of the core collapse
time: $t_d \approx 2.5\, t_{cc}$.

\begin{figure} \plotone{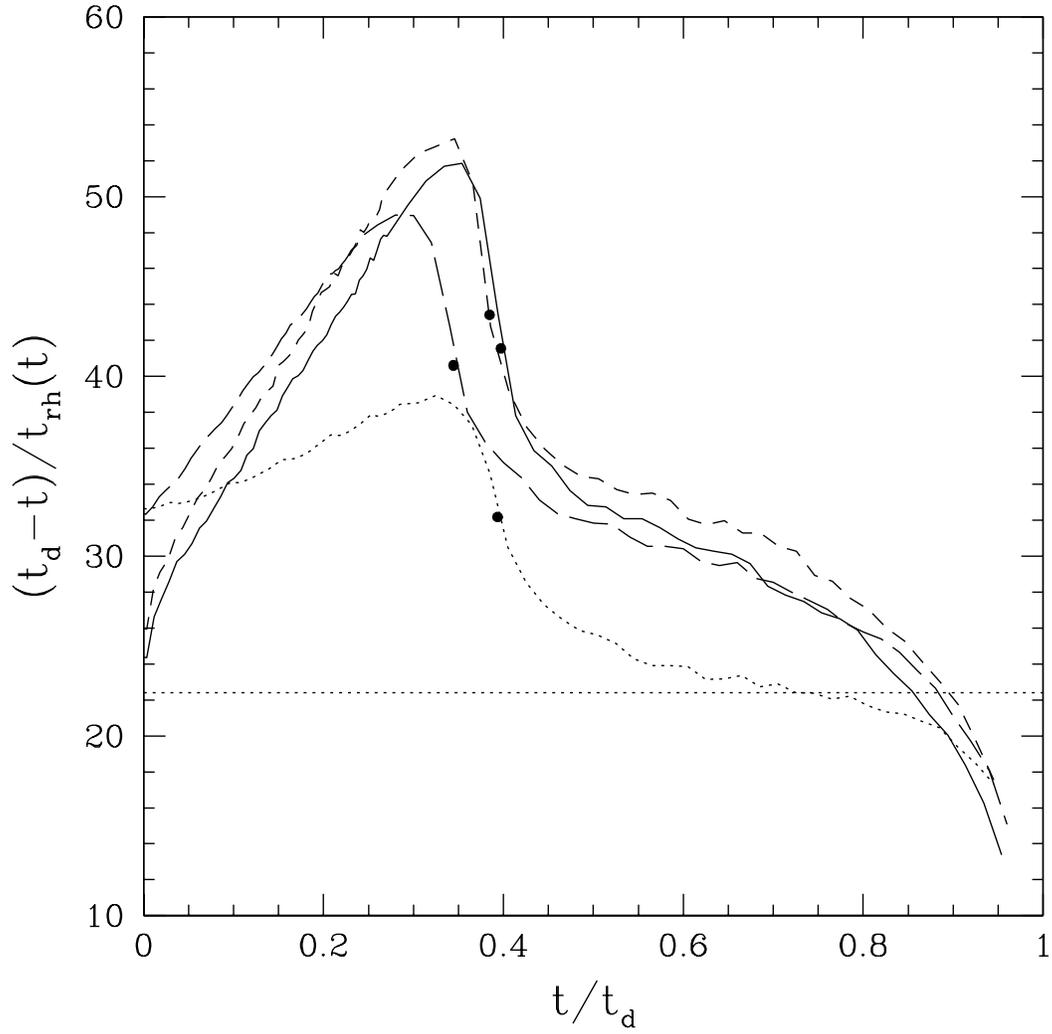}
\figcaption{Time to destruction of the cluster \ngc\ in units of the
  current relaxation time, $t_{rh}(t)$, vs. cluster age.  Line notation
  is the same as in Fig. \protect\ref{fig:m}.  The filled circles
  indicate the time of maximum core collapse.  The dotted horizontal
  line is H\'{e}non's self-similar solution, equation
  (\protect\ref{eq:Henon}).  The abscissa is normalized to the different
  destruction time $t_{d}$ for each model.  \label{fig:tremain}}
\end{figure}

The post-collapse evolution of globular clusters depends on the
concentration and the remaining mass.  The self-similar solution by
\cite{H:61} (1961) for tidally-limited clusters can be expressed
(Goodman 1993) as follows:
\begin{equation}
t_{d}-t = 22.4 \, t_{rh}(t),
\label{eq:Henon}
\end{equation}
where $t_{d}-t$ is the time to disruption at time $t$.  Figure
\ref{fig:tremain} shows that the self-similar limit is approached at
late stages of the cluster evolution, but for the most part equation
(\ref{eq:Henon}) underestimates the time to destruction by about 50\%.

\subsubsection{Comparison of adiabatic corrections}

An important constituent of our models is adiabatic corrections for the
tidal shock effects (eqs. \ref{eq:SpCorr}--\ref{eq:NumCorr}).  Various
corrections predict different impact of shocks in the middle parts of
clusters and lead to different evolutionary paths.  The amplitude of the
corrections depend on the ratio of the perturbation timescale to the
dynamical time of stars in the cluster.  For \ngc, the characteristic
time for disk shocking (eq. [\ref{eq:tau}]) is $\tau_{\rm disk} =
1.3\times 10^6$ yr, and for bulge shocking (cf. \cite{GHO:99} 1999) is
$\tau_{\rm bulge} = 1.3\times 10^7$ yr.  The half-mass dynamical time of
the cluster is initially much shorter, $\omega_h^{-1} = 3.1\times 10^5$
yr, leading to strong suppression of the effects of tidal shocks.

\begin{figure} \plotone{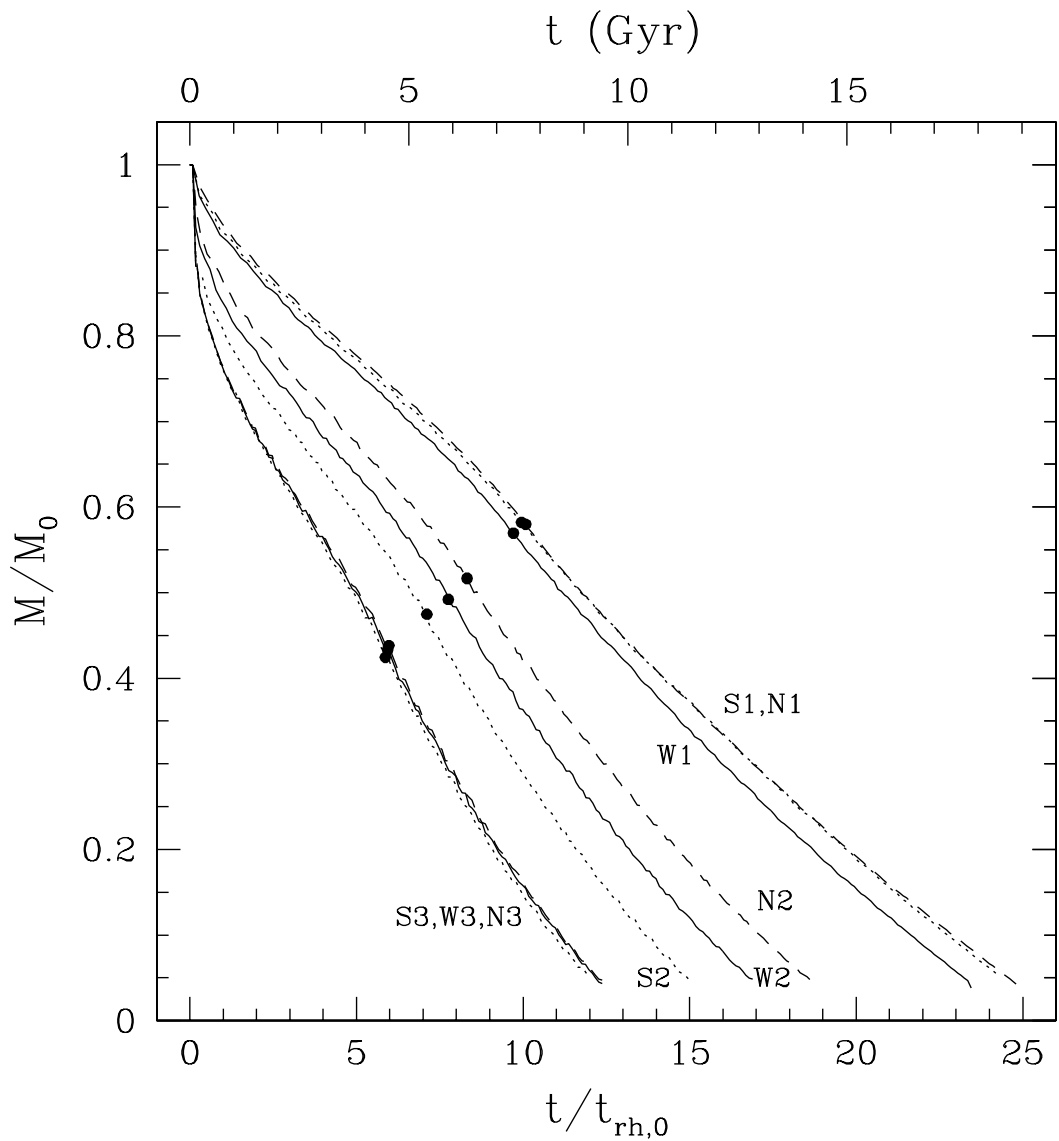}
\figcaption{Comparison of the Spitzer ({\it S}), Weinberg ({\it W}), and
  \N-body ({\it N}) adiabatic corrections for the {\it case 3} models of
  \ngc.  The first set of models ({\it S1,W1,N1}) is calculated with the
  true values of the disk and bulge shock timescales, $\tau_{\rm disk}$
  and $\tau_{\rm bulge}$; the second set ({\it S2,W2,N2}) with the shock
  timescales reduced by a factor of 5; and the third set ({\it
  S3,W3,N3}) with the shock timescales reduced by a factor of 100.
  Solid lines correspond to the Weinberg corrections
  (eq. [\protect\ref{eq:WCorr}]), dots to the Spitzer corrections
  (eq. [\protect\ref{eq:SpCorr}]), and dashes to the \N-body corrections
  for fast shocks (eq. [\protect\ref{eq:NumCorr}]).  Filled circles
  indicate the time of maximum core collapse.  \label{fig:mad}}
\end{figure}

Figure \ref{fig:mad} compares the Spitzer, Weinberg, and \N-body
adiabatic corrections for the {\it case 3} models of \ngc.  When the
shocks are ``slow'' ($x = \omega_h \tau \gg 1$), as indicated by the
true values of $\tau_{\rm disk}$ and $\tau_{\rm bulge}$, both {\it S1}
and {\it N1} models have tidal shock effects suppressed more strongly
than does the {\it W1} model with the less steep adiabatic corrections.
For the ``slow'' shocks, the {\it W1} model gives the correct
description of the cluster evolution.

When we decrease the shock timescales by a factor of five, we enter the
regime of ``fast shocks'' and the {\it N2} model is correct.  For $x
\lesssim 1$, we can estimate the importance of the adiabatic corrections
by the Taylor expansion of equations
(\ref{eq:SpCorr}--\ref{eq:NumCorr}).  The {\it N2} model shows the
slowest evolution because $A_{\rm 1N} \approx 1 - 2.5 x^2$, whereas for
the {\it W2} model, $A_{\rm 1W} \approx 1 - 1.5 x^2$.  Even though the
first order correction for the {\it S2} model, $A_{\rm 1S} \approx 1 - 2
x^2$, is smaller than $A_{\rm 1W}$, the second order correction is
larger, $A_{\rm 2S} \approx 1 - 1.1 x^2$.  This allows the {\it S2}
model to evolve faster than the other two.

Finally, when we reduce the shock timescales by a factor of hundred, the
impulse approximation applies everywhere in the cluster and all three
models are essentially identical.  The evolution in this case proceeds
much faster: the cluster is destroyed in half the time of the original
model {\it W1}.  Thus, adiabatic corrections are critical for
calculating correct globular cluster models.

\subsubsection{Low concentration model}

For comparison, we consider another model of \ngc, with lower
concentration $c=0.84$, or $W_0=4$.  This is the model used for \N-body
simulations of \cite{GO:99} (1999).  All other parameters are the same
as in the previous model.  We keep the observed tidal radius in parsecs,
so the core radius and the relaxation time scale in physical units
accordingly (see Table \ref{tab:param}).  Therefore, both models have
the same average density, $\rho_{\rm av} \propto M/R_t^3$.

\begin{figure} \plotone{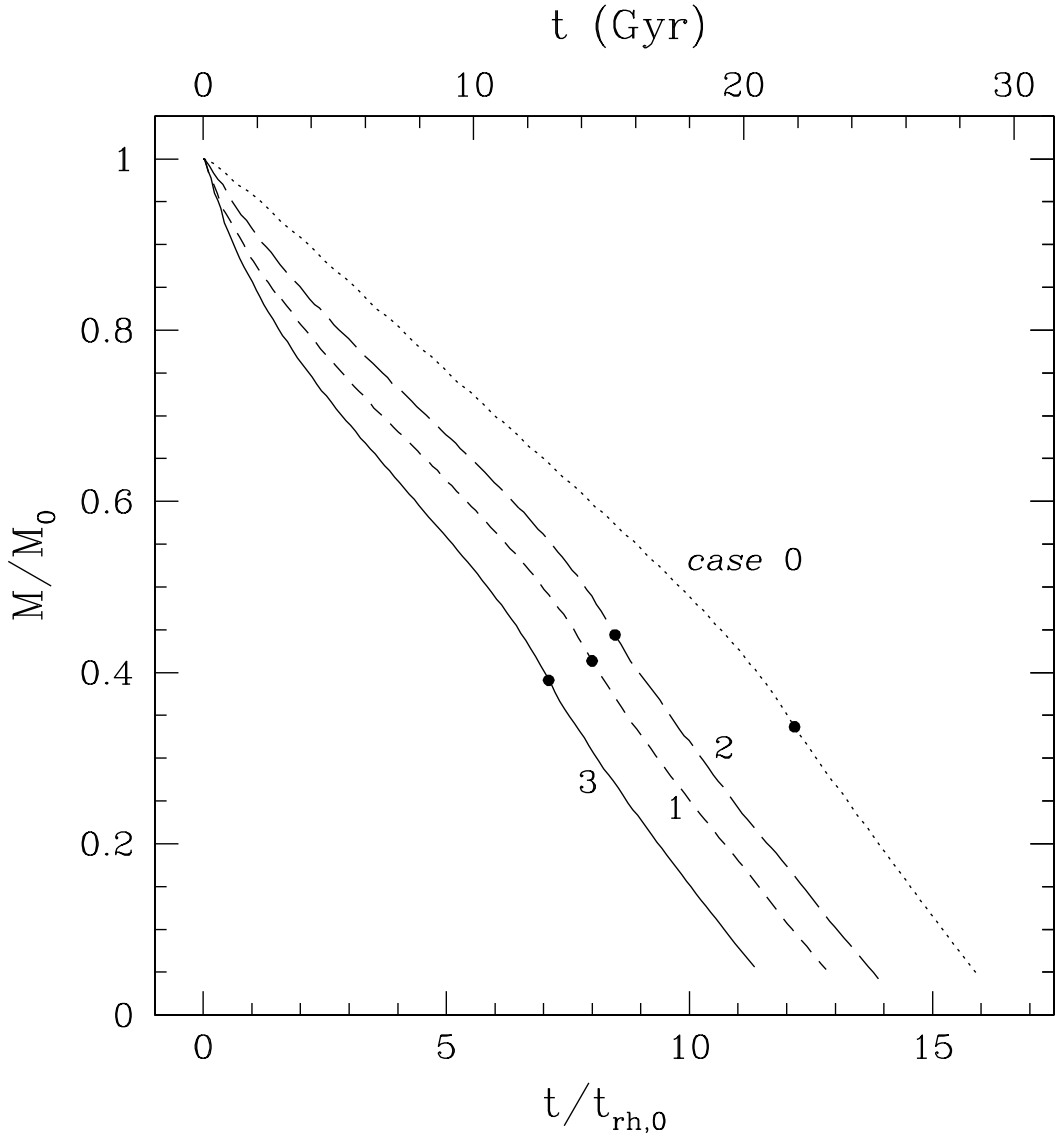}
\figcaption{The mass loss of the low concentration model, c=0.84.  All
  other parameters of \ngc\ are fixed, including the tidal radius in
  parsecs.  Line notation is the same as in Fig.  \protect\ref{fig:m}.
  \label{fig:m1}}
\end{figure}

\begin{figure} \plotone{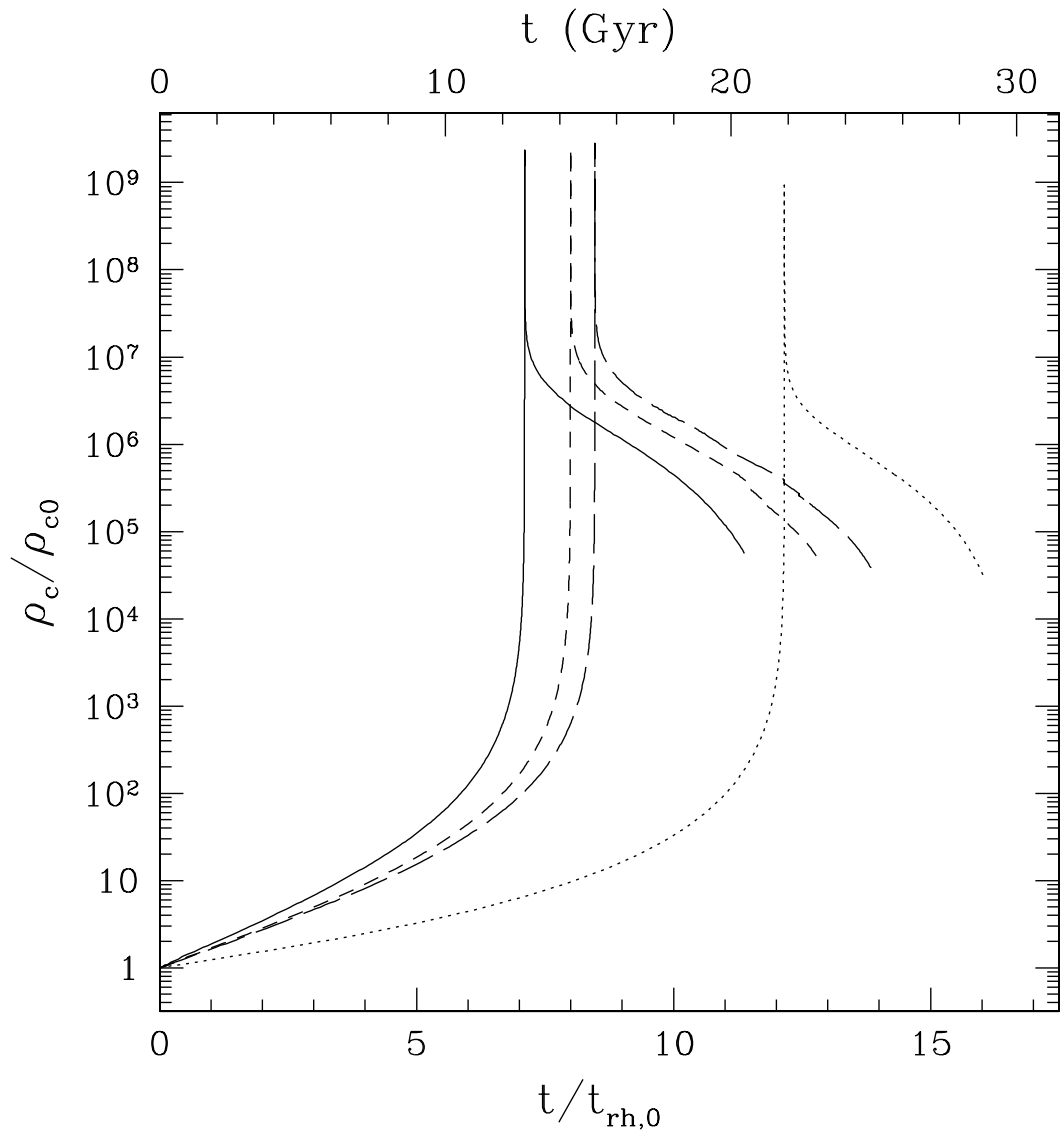}
\figcaption{Central density of the low concentration cluster.  Line
  notation is the same as in Fig. \protect\ref{fig:m}.
  \label{fig:rho1}}
\end{figure}

Figure \ref{fig:m1} shows the mass evolution of the low concentration
model.  The cluster evaporates much faster than the high concentration
model, in units of the initial relaxation time $t_{rh,0}$.  Clearly,
tidal shocks have much stronger effect on less concentrated clusters but
the total speed-up in time to destruction, due to shocks, is again about
25\%.  Notice that the destruction time in years is about the same for
the low- and high-concentration models.  Figure \ref{fig:rho1} shows
that the evolution of the central density is qualitatively similar to
the previous model.  However, the effect of the tidal shock relaxation
is stronger for the low-concentration model.

\subsection{Multi-mass models}

In a single mass model, the energy is transferred between the cluster
core and the envelope, which requires a conduction mechanism acting on
the half-mass relaxation timescale.  In a multi-mass model, the energy
exchange occurs also between stars of different mass.  A typical
timescale for the energy exchange between the two mass species $m_1$ and
$m_2$ (where $m_2<m_1$) is $t_{rh} \times (m_2/m_1)$.  Since this is
much shorter than the core collapse time for single mass clusters, the
overall evolution of multi-mass clusters is expected to be faster.

\cite{LG:95} (1995), for example, considered the evaporation of a
multi-mass cluster in a steady tidal field and found that the mass loss
rate can more than double relative to the single-component case.
\cite{CW:90} investigated the effects of stellar evolution on the
cluster dynamics.  In the future, it will be important to generalize the
present calculations by including a realistic mass function and stellar
evolution, along with the presently implemented two-body relaxation and
tidal shocks.  In this paper, we present only an illustrative model
allowing for the mass spectrum.

We take the previously described model for \ngc\ and include seven mass
components, ranging from $0.1 M_{\sun}$ to $0.7 M_{\sun}$.  The number
of stars in each component $N(m) \propto m^{-2}$.  Thus, the upper mass
limit corresponds to the turn-off of the main sequence at 10 billion
years (we also used it in the single-mass models), and the lower limit
is a reasonable cutoff of the initial mass function.  The relaxation
time is defined by equation (\ref{eq:trh}), where we substitute for
$m_*$ the mean stellar mass (see \cite{LG:95} 1995).

\begin{figure} \plotone{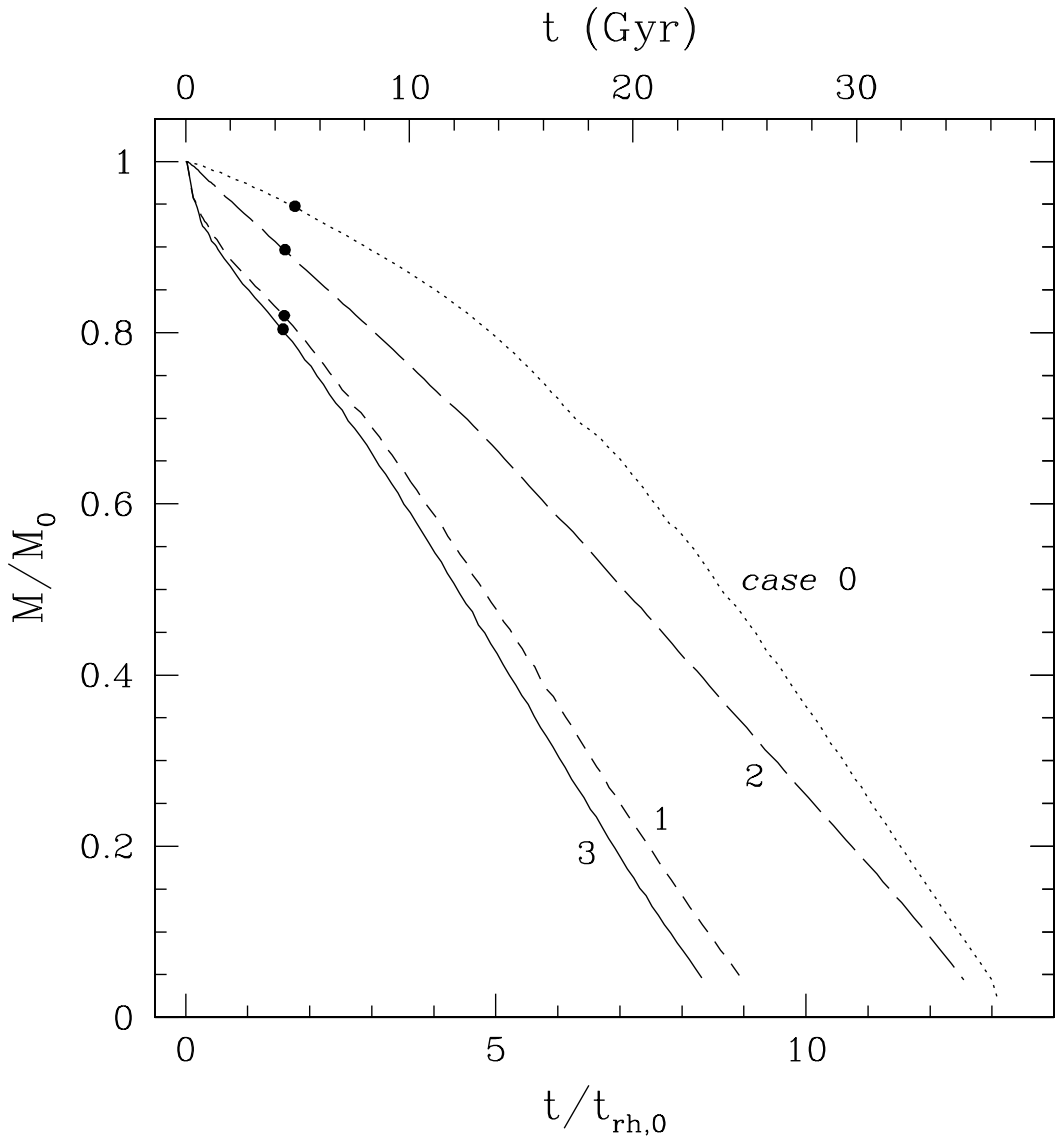}
\figcaption{The mass loss of the multi-mass model of \ngc.  Line
  notation is the same as in Fig. \protect\ref{fig:m}.
  \label{fig:m_multi}}
\end{figure}

Figure \ref{fig:m_multi} shows the mass loss for the multi-mass model.
Notice that although the evolution proceeds faster in units of the
initial half-mass relaxation time, this timescale expressed in years is
longer than that for the single-component model (Table \ref{tab:param}).
As a result, the destruction time is roughly 20 Gyr in both cases.  The
contribution of tidal shocks in the multi-mass model is greater,
reducing the time to destruction by about 37\%.

\subsubsection{Comparison with \N-body models}

Recent paper by \cite{TPZ:98} (1998) compares the multi-mass
two-dimensional Fokker-Planck (F-P) models of \cite{T} (1995) with the
\N-body simulations using a special purpose computer, {\it GRAPE-4}.
The models include stellar evolution, two-body relaxation, and the tidal
truncation for a low concentration ($c = 0.67$) cluster model from
\cite{CW:90}.  The above authors argue that the escape of stars through
the tidal boundary in an isotropic F-P code, where the distribution
function depends only on the energy, is faster than in the \N-body
models.  To remedy this disagreement, they suggest a new escape
criterion based both on the energy and angular momentum of stars.  This
``apocenter'' criterion can only be implemented in the anisotropic F-P
code of \cite{Tetal:97}.

The old isotropic and the new anisotropic escape criteria lead to
significantly different results only when the mass loss rate per
relaxation time is high.  This especially affects the early stages of
the cluster evolution, when stellar winds from young massive stars drive
mass loss on a relatively short timescale.  Immediate removal of that
mass in the \cite{CW:90} models lead to the disagreement with the
\N-body simulations (see \cite{TPZ:98} 1998).  Since our models do not
include stellar evolution, we expect less disagreement.

We run a {\it case 0} model of the cluster using the parameters from
\cite{CW:90}.  We use 14 mass components from $0.4 M_{\sun}$ to $15
M_{\sun}$, distributed according to the specified mass function, $N(m)
\propto m^{-2.5}$.  Without stellar evolution the time to destruction in
our model should be longer than in the \N-body and the anisotropic F-P
models which give $t_d \approx 1.3 t_{rh,0}$.  Indeed, we find that our
model evolves slower, with $t_d \approx 2 t_{rh,0}$.  This indicates
that our results represent an upper limit on the destruction times of
globular clusters.


\section{Review of Globular Cluster Evolution Including Tidal Shocks}
\label{sec:review}

The evolution of the isolated single-mass King models (in dimensionless
units) is determined by the only parameter, the concentration $c$.
Inclusion of tidal shocks introduces another independent parameter, for
example, the ratio of the half-mass relaxation time to the
characteristic shocking time,
\begin{equation}
\beta \equiv t_{rh}/t_{sh}.
  \label{eq:betadef}
\end{equation}
The parameter $\beta$ is a combination of two variables describing the
structure of clusters: the number of stars, $N$, and the characteristic
density, $\rho_h \propto M/R_h^3$.  When the amplitude of the tidal
force on the cluster is fixed, the shocking time (eq. [\ref{eq:tsh}])
scales as $t_{sh} \propto \rho_h$, and the relaxation time
(eq. [\ref{eq:trh}]) scales as $t_{rh} \propto \rho_h^{-1/2}
N/\ln{\Lambda}$, where $\Lambda=0.4N$.  Then
\begin{equation}
\beta \propto {N \over \ln{\Lambda}} \; \rho_h^{-3/2}.
  \label{eq:beta}
\end{equation}
While the concentration $c$ describes the effects of two-body
relaxation, the parameter $\beta$ shows the relative importance of tidal
shocks.  In this Section, we explore the evolution of globular clusters
for a wide range of initial parameters $c$ and $\beta$.

All models in this section include fully the relaxation and tidal shock
effects (i.e., {\it case 3}) with the adiabatic corrections for ``slow
shocks'', equation (\ref{eq:WCorr}).  We arrange a grid of initial
conditions, covering a number of cluster concentrations, from $c=0.6$ to
$c=2.6$ with the step 0.2, and a range of shock parameters, $\beta$.  We
compute 15 models per concentration family, equally spaced on the
logarithmic scale from $\beta \approx 10^{-5}$ to $\beta \approx 10^2$.
All other parameters correspond to the model of \ngc\ described in the
previous section.  The observed tidal radius of \ngc\ is fixed and the
core radius scales accordingly with the concentration.  Thus all models
have initially the same average density, $M/R_t^3$.

\subsection{Cluster evolution in figures}

\begin{figure} \plotone{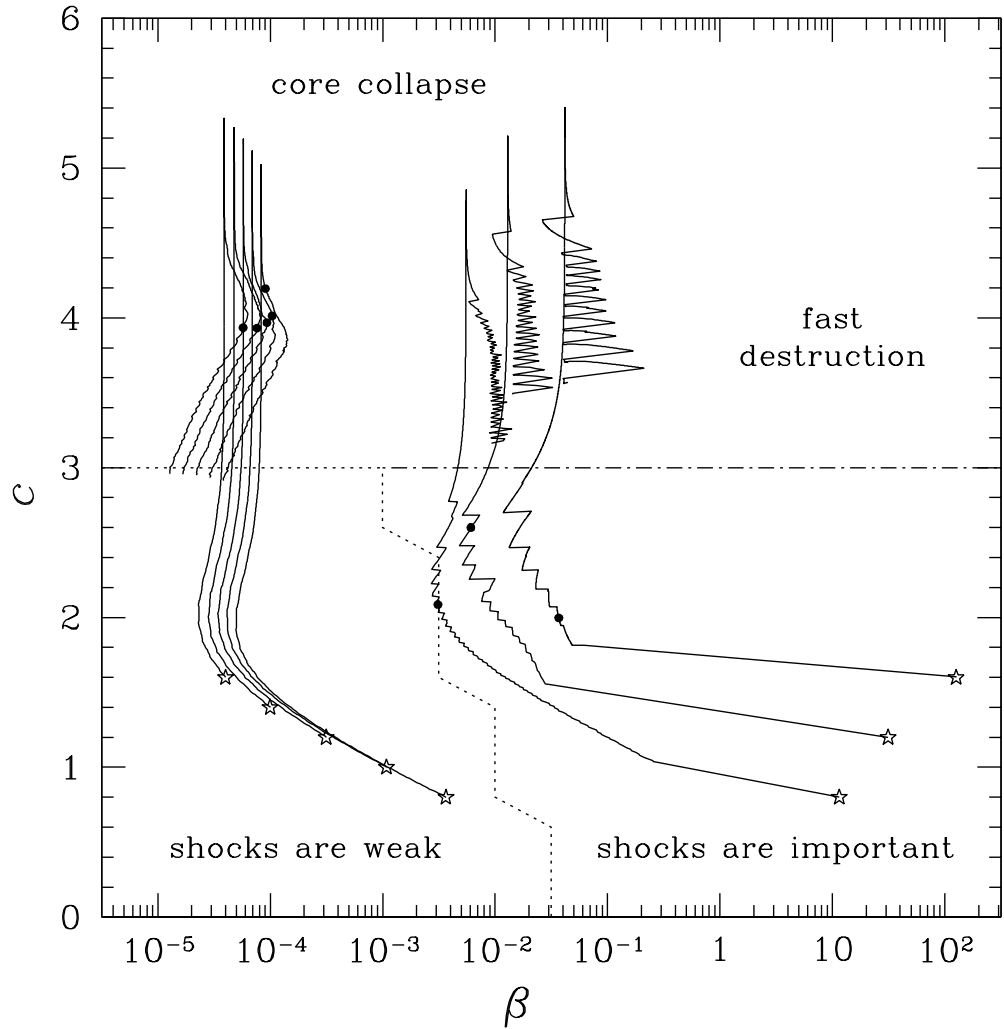}
\figcaption{Evolution of the model clusters as a function of the
  concentration, $c$, and the shock parameter, $\beta$.  Stars show the
  initial conditions, filled circles the time of core collapse.  Marked
  regions on the diagram correspond to the different regimes of cluster
  evolution.  The critical values of $\beta$, separating the left and
  right regions, are determined when the core collapse time deviates
  from its value for $\beta=0$.  \label{fig:cbeta}}
\end{figure}

Figure \ref{fig:cbeta} shows how different regimes of cluster evolution
map on the plane of parameters $c$ and $\beta$.  In the lower left
region of the plot tidal shocks are weak and evolution is dominated by
two-body relaxation; in the upper left region core collapse proceeds on
a relatively short timescale without affecting most of the cluster; and
in the lower right region tidal shocks are strong and cause substantial
mass loss before core collapse.  The upper right region is never reached
in reality because of extremely strong tidal shocks leading to fast
destruction of the clusters.

\begin{figure} \plotone{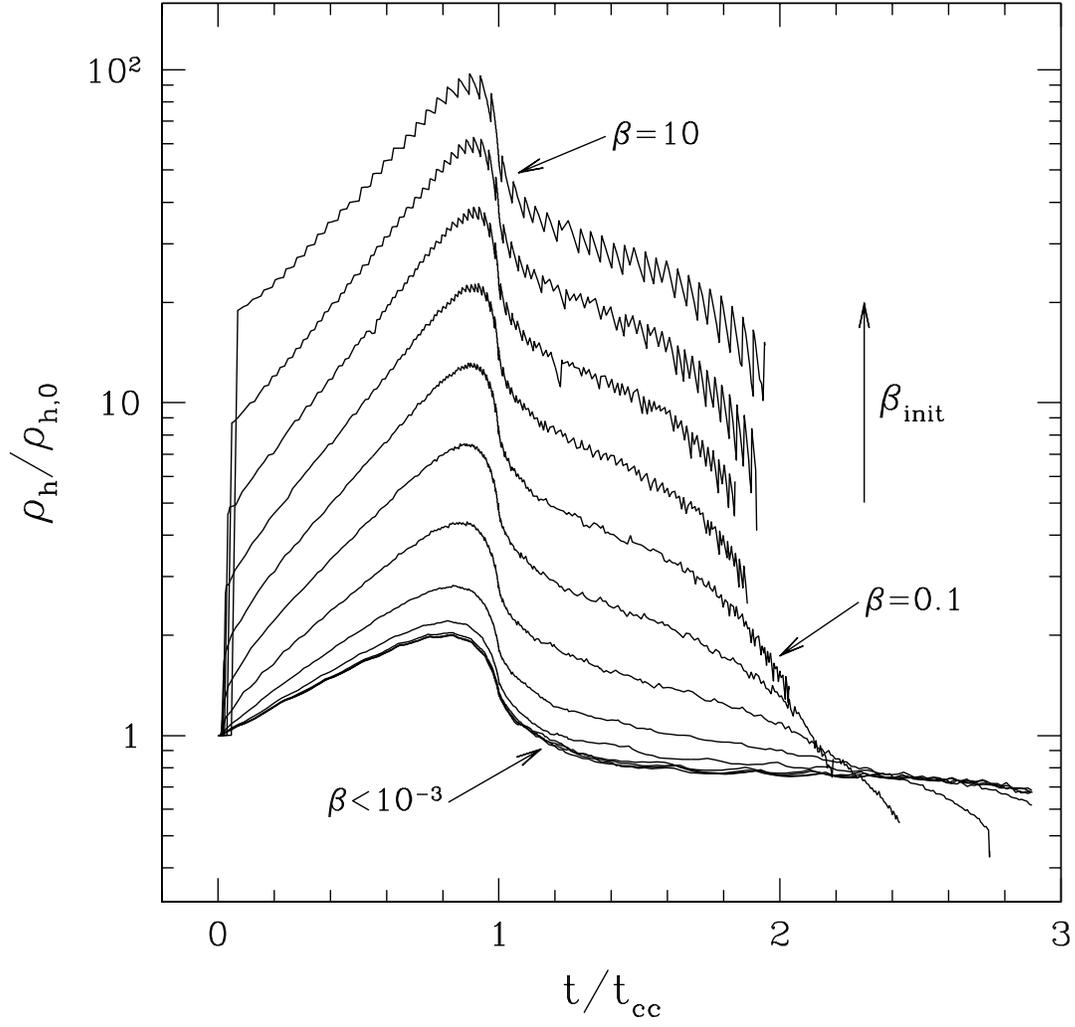}
\figcaption{The mean density, $\rho_h$, as a function of time normalized
  to the individual core collapse time for each run.  All models shown
  have the same initial concentration $c=1.4$ but different shock
  parameters varying from $\beta=10^{-5}$ to $\beta=10$.  Models with
  $\beta < 10^{-3}$ show essentially identical evolution.  At the time
  of core collapse, the mean density is higher for the higher values of
  $\beta$.  \label{fig:rhomean}}
\end{figure}

Qualitative evolution in the left region of the plane is characteristic
of a tidally-truncated model.  As clusters start to collapse, the
relaxation time becomes increasingly shorter and the tidal shock time
becomes increasingly longer because the mean density $\rho_h$ rises
(Figure \ref{fig:rhomean}).  The mass loss is weak at this stage and the
number of stars, $N$, stays relatively constant.  As a result, the
parameter $\beta$ decreases (and moves to the left of the diagram) until
the very advanced stages of core collapse.  At that point the value of
$\beta$ freezes and the evolutionary tracks are strictly vertical (the
upper part of the diagram).  After core collapse, the cluster re-expands
and its density falls slightly, moving it to the right of the $c-\beta$
diagram.  At a later time, so little mass is left that two-body
relaxation speeds up again and drives clusters to complete dissolution.

\begin{figure} \plotone{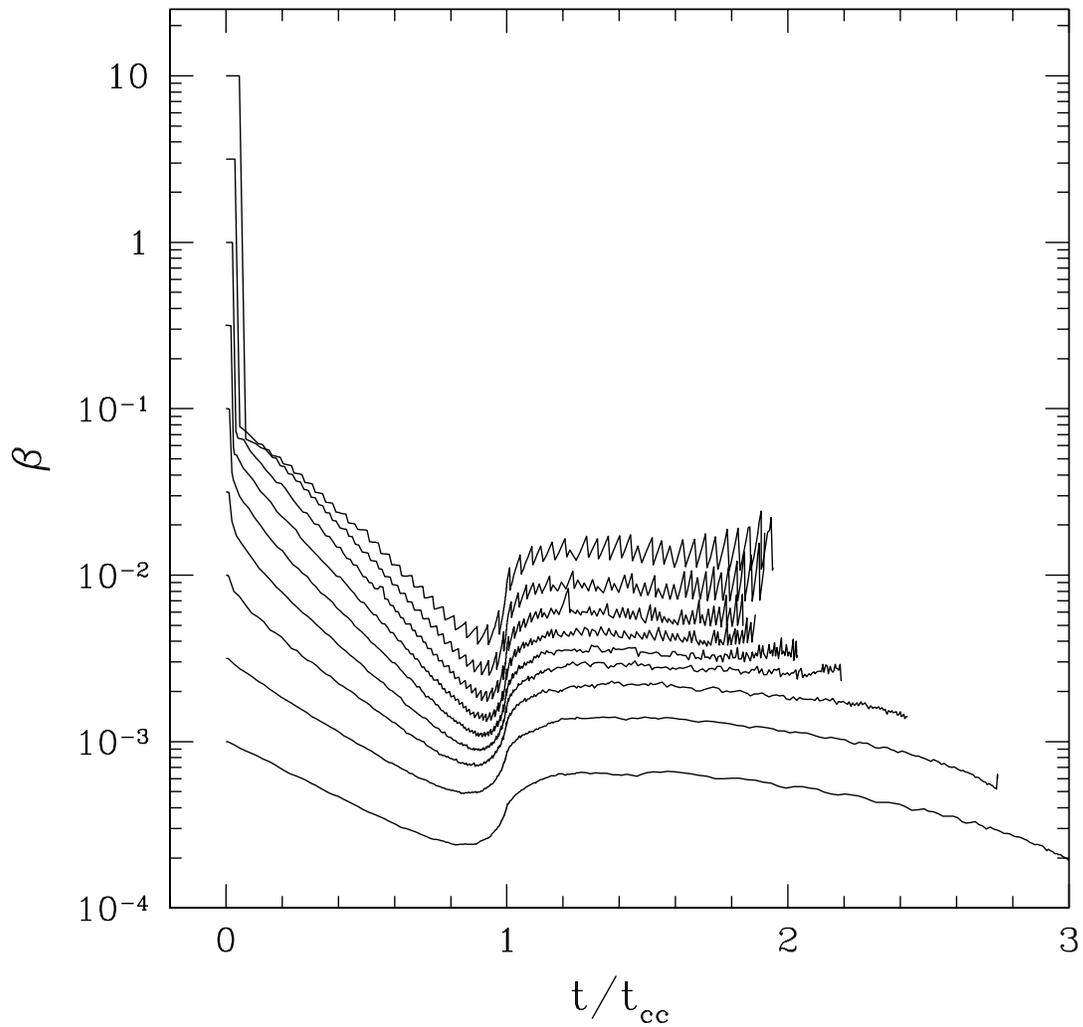}
\figcaption{The shock parameter, $\beta$, as a function of time
  normalized to the individual core collapse time, for the models with
  the various initial values of $\beta$ and the same initial
  concentration $c=1.4$.  \label{fig:beta}}
\end{figure}

Figure \ref{fig:beta} illustrates the behavior of the parameter $\beta$
with time.  On the low end, all lines scale similarly, following the
evolution of the mean density.  For the large initial values of $\beta$,
the early mass loss due to shocks is very important.  This leads us to
investigate the right part of the $c-\beta$ diagram.

The evolution of clusters with strong tidal shocks differs dramatically
from the previous case.  An early mass loss caused by the shocks changes
the structure of the clusters, removing stars from the outer parts and
adding energy dispersion in the core.  This increases the central and
the mean density of the clusters.  Both core collapse and final
destruction proceed much faster.  These effects of tidal shocks leave
noticeable ripples in the clusters tracks on the right part of the
diagram (Figure \ref{fig:cbeta}).

Tidal shocking is rapidly self-limiting.  Clusters with large values of
$\beta$ quickly lose mass and evolve to $\beta \lesssim 0.1$.  Figure
\ref{fig:beta} demonstrates that just a first few shocks lower the value
of $\beta$ by several orders of magnitude.  Every subsequent shock
causes the see-saw variations of the shock parameter, with an increasing
amplitude towards the late stage of evolution when fewer stars are left.

A few clusters in the right part of the $c-\beta$ diagram cross their
evolutionary tracks on the way to core collapse.  At the time of
crossing, at least one of the clusters has already suffered a severe
mass loss caused by the shocks and has a structure significantly
deviating from King models.  The density profiles of the two clusters
are similar within the half-mass radii but depart from each other closer
to the tidal radius.  Those clusters with higher initial concentration
and stronger shocks are more severely truncated than the clusters with
smaller concentration and weaker shocks.  Also, tidal shocks add the
velocity dispersion in the core and, therefore, raise the central
density so that the ratio $R_t/R_c$, as measured by the concentration
$c$, is the same for both clusters.

At the crossing point the two parameters, $c$ and $\beta$, no longer
uniquely specify cluster's future path.  A third parameter comes into
play as we investigate the thermodynamics of clusters.

\subsection{Cluster thermodynamics}

From a thermodynamic point of view, we can consider stars as particles
of gas.  A critical factor determining evolution is the heat flow
between the core and the halo of the cluster.  The heat in this case is
the velocity dispersion of stars, $T \equiv v_m^2$, whose flux is
derived in \cite{S:87} (1987, p. 68).  The amount of heat transported
through the cluster per unit time, or the ``luminosity'', is $4\pi r^2
F(r)$.  The heat conduction facilitates core collapse and proceeds on a
relaxation timescale.  Let us construct a new dimensionless ratio of the
heat transported through the half-mass radius in the relaxation time
$t_{rh}$, to the total reservoir of heat, $T(r_h)$.  Neglecting constant
factors of order unity, we define the following parameter
\begin{equation}
\gamma \equiv - \left.{d\ln{v^2} \over d\ln{r}}\right|_{r_h},
\end{equation}
which is essentially a logarithmic gradient of the velocity dispersion.
For an isothermal sphere, $\gamma = 0$.  In general, for clusters on the
way to core collapse, this parameter is positive as the heat is
transferred from the kinematically hot core to the cold halo and grows
as core collapse accelerates.  In contrast, tidal shocks try to reverse
the heat flow by heating preferentially the outer parts of the cluster.
Even though the instantaneous value of $\gamma$ is determined only by
the density structure, its derivative, $\dot{\gamma}$, is a signature of
the relaxation-- or the tidal shock--driven evolution.

The clusters crossing paths on the $c - \beta$ diagram have similar
values of $\gamma$, but the derivatives have opposite sign.  For the
clusters moving almost vertically on the diagram $\dot{\gamma}>0$,
whereas for the clusters which cross the diagram horizontally
$\dot{\gamma}<0$.  Therefore, the evolution of the former is dominated
by two-body relaxation and of the latter by tidal shocks.

Knowing the value of $\dot{\gamma}$ for the Galactic globular clusters,
we could in principle differentiate the two evolutionary paths.
Unfortunately, it is hard to establish the sign of $\dot{\gamma}$
observationally.

Another signature of the tidal shock--dominated clusters is a steep
density profile and a low number density of stars at the tidal radius.
Strong shocks sweep away a large number of stars at once and leave the
cluster filling smaller volume in space than the Roche lobe imposed by
the external tidal field.  Clusters with a sharp density contrast and a
high velocity dispersion of the halo stars might be experiencing strong
shocking.  Most of such clusters are expected to come close to the
Galactic center but are not necessarily present there now.

\subsection{Cluster evolution in numbers}

\begin{figure} \plotone{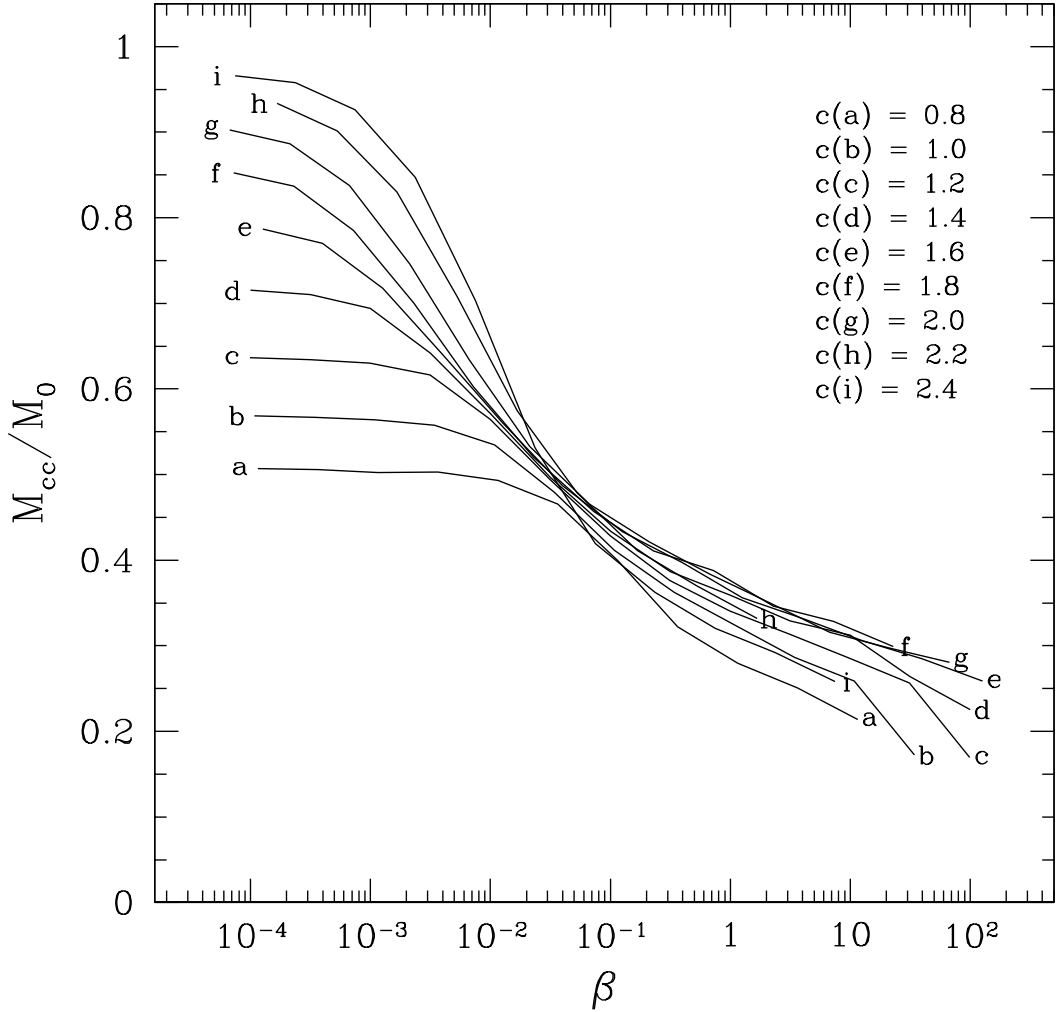}
\figcaption{The mass remaining at the time of core collapse vs. the
  shock parameter, $\beta$, for the various families of cluster models.
  Each family has the same initial concentration, from $c=0.8$ for
  family {\it a} to $c=2.4$ for family {\it i}.  Letters correspond to
  the end models of each family.  \label{fig:mcc}}
\end{figure}

The mass loss from clusters is strongly enhanced by tidal shocks.  An
illustration of this is the fraction of the initial mass remaining at
the time of core collapse, $M(t_{cc})$.  Figure \ref{fig:mcc} shows that
the larger $\beta$, the smaller the remaining mass.  Note, that many
models of various initial concentration converge at $\beta \approx 0.1$,
giving $M(t_{cc}) \approx 45\%$.

The core collapse time is a strong function of both initial parameters
$c$ and $\beta$.  Figure \ref{fig:tcc_rel} shows $t_{cc}$ for the models
without tidal shocks.  In units of the initial half-mass relaxation
time, the core collapse time can be fitted as follows:
\begin{equation}
{t_{cc} \over t_{rh,0}} = f_1(c) \equiv
  10^{a_1 + a_2 c + a_3 c^2 + a_4 c^3 +a_5 c^4},
  \label{eq:tcc_rel}
\end{equation}
where the coefficients $a_i$ are given in Table \ref{tab:fitc}.  Our
results are in good agreement with \cite{Q:96}, except for the clusters
with very high concentration.  Even for them, the maximum difference in
$t_{cc}$ is less than 30\%.

\begin{figure} \plotone{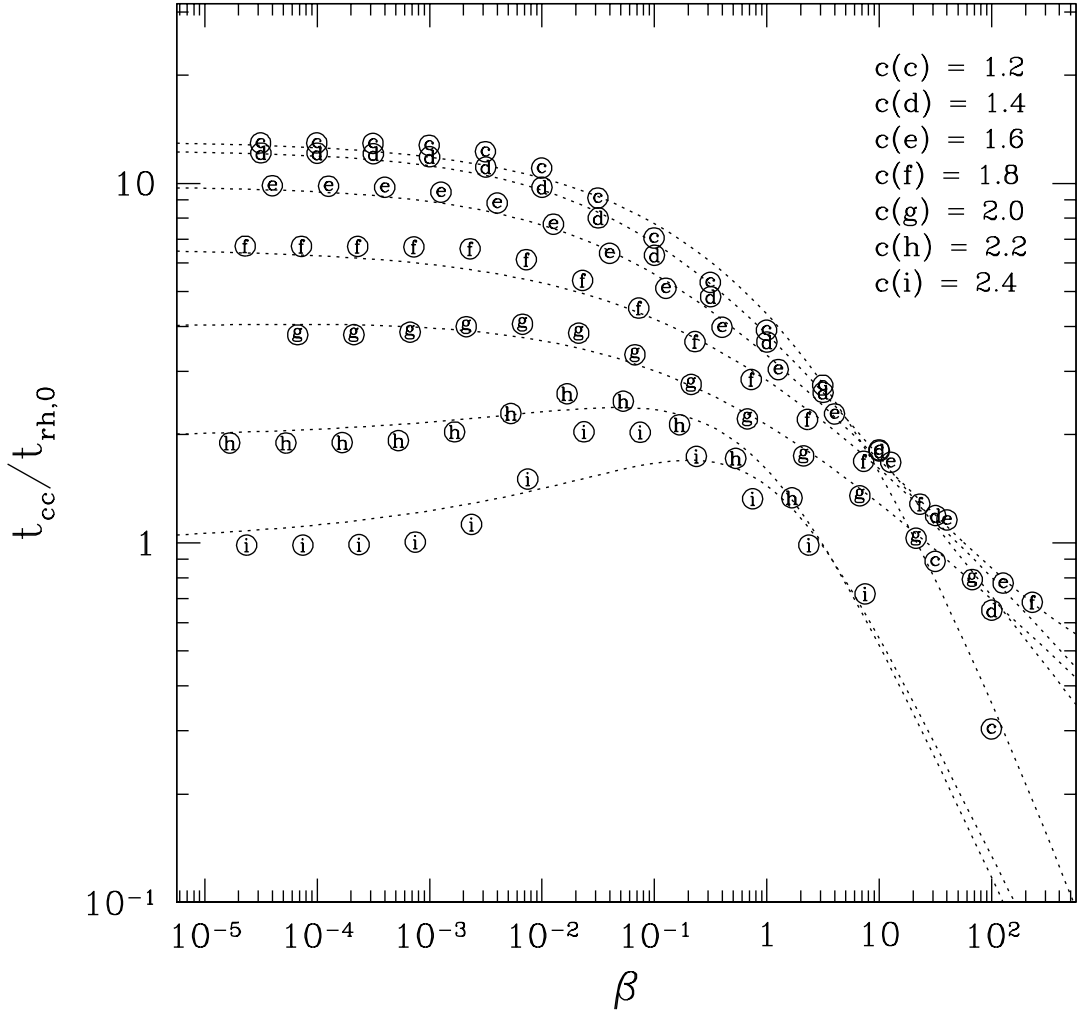}
\figcaption{The core collapse time as a function of the shock parameter,
  $\beta$.  The families of models with the initial concentrations
  $c=0.8$ ({\it a}) and $c=1.0$ ({\it b}) are excluded for clarity.
  Dotted lines are our fit, eq. (\protect\ref{eq:tcc_sh}).
  \label{fig:tcc_sh}}
\end{figure}

Figure \ref{fig:tcc_sh} shows the core collapse time for the models
including tidal shocks.  The shock--induced relaxation speeds up core
collapse by adding velocity dispersion in the core and by removing stars
in the outer parts of the clusters, thus reducing the relaxation time.
We fit the results assuming, as an {\it Ansatz}, the following
functional form:
\begin{equation}
{t_{rh,0}\over t_{cc}} = \left. {t_{rh,0}\over t_{cc}}\right|_{\beta=0} \;
  \left( 1 + b_1 \beta^{b_2} + b_3 \beta^{b_4} \right).
  \label{eq:tcc_sh}
\end{equation}
The coefficients $b_i$ for models of various concentration $c$ are given
in Table \ref{tab:fittccbeta}.  The fit is generally accurate to 20\%,
with larger errors for models with very high initial concentration.

\begin{figure} \plotone{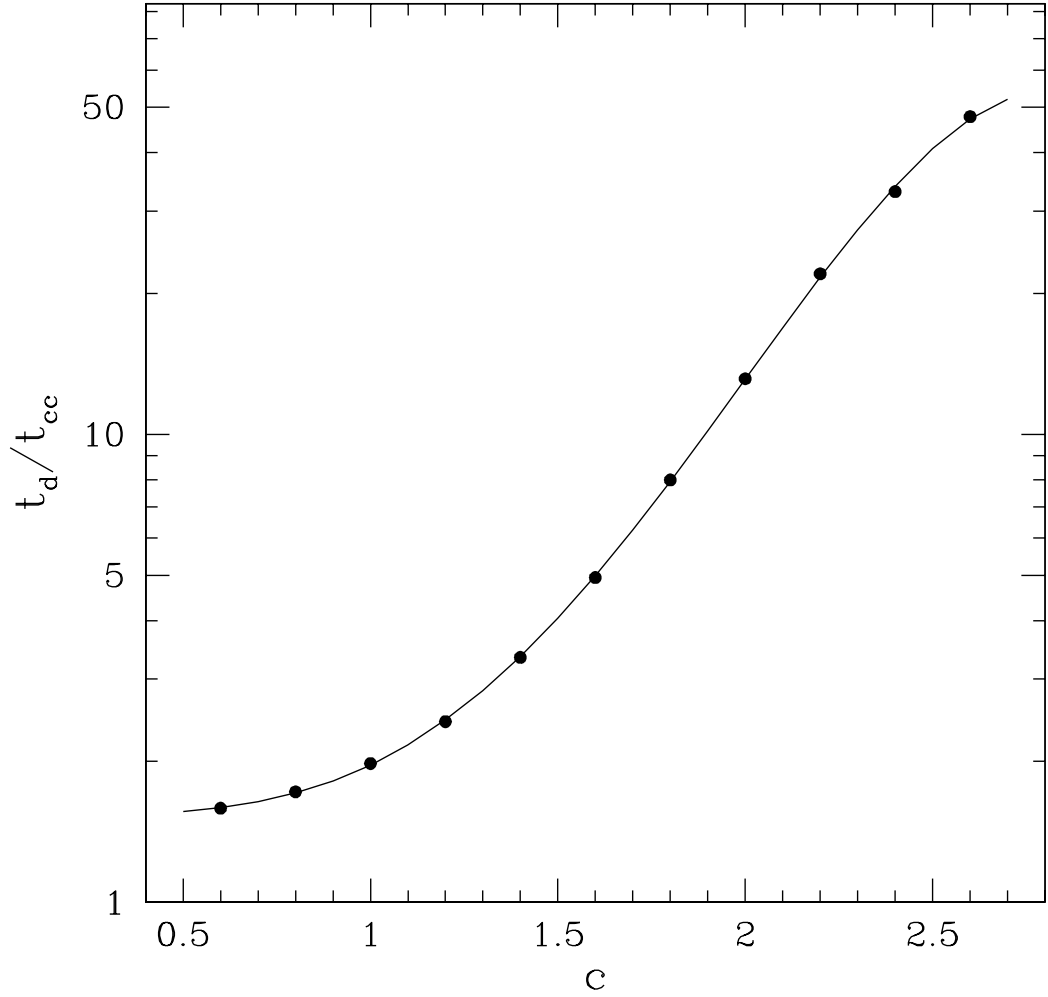}
\figcaption{The destruction time in units of the core collapse time
  vs. cluster concentration for the models without tidal shocks.  The
  solid line is our fit, eq. (\protect\ref{eq:td_rel}).
  \label{fig:td_cc}}
\end{figure}

The time to destruction, $t_d$, is also strongly affected by tidal
shocks.  As we have emphasized in the previous section, evolution of
clusters with different parameters scales similarly when time is
normalized to the core collapse time.  Accordingly, we provide the
fitting formulae for the scaled destruction time, $t_d/t_{cc}$.  For the
models without tidal shocks (Figure \ref{fig:td_cc}), we use the same
functional form as equation (\ref{eq:tcc_rel}) and fit $t_d$ both in
units of the initial relaxation time and in units of the core collapse
time:
\begin{equation}
{t_d \over t_{rh,0}} = f_2(c), \hspace{1cm}
{t_d \over t_{cc}} = f_3(c),
  \label{eq:td_rel}
\end{equation}
where the coefficients $a_i$ are again given in Table \ref{tab:fitc}.

\begin{figure} \plotone{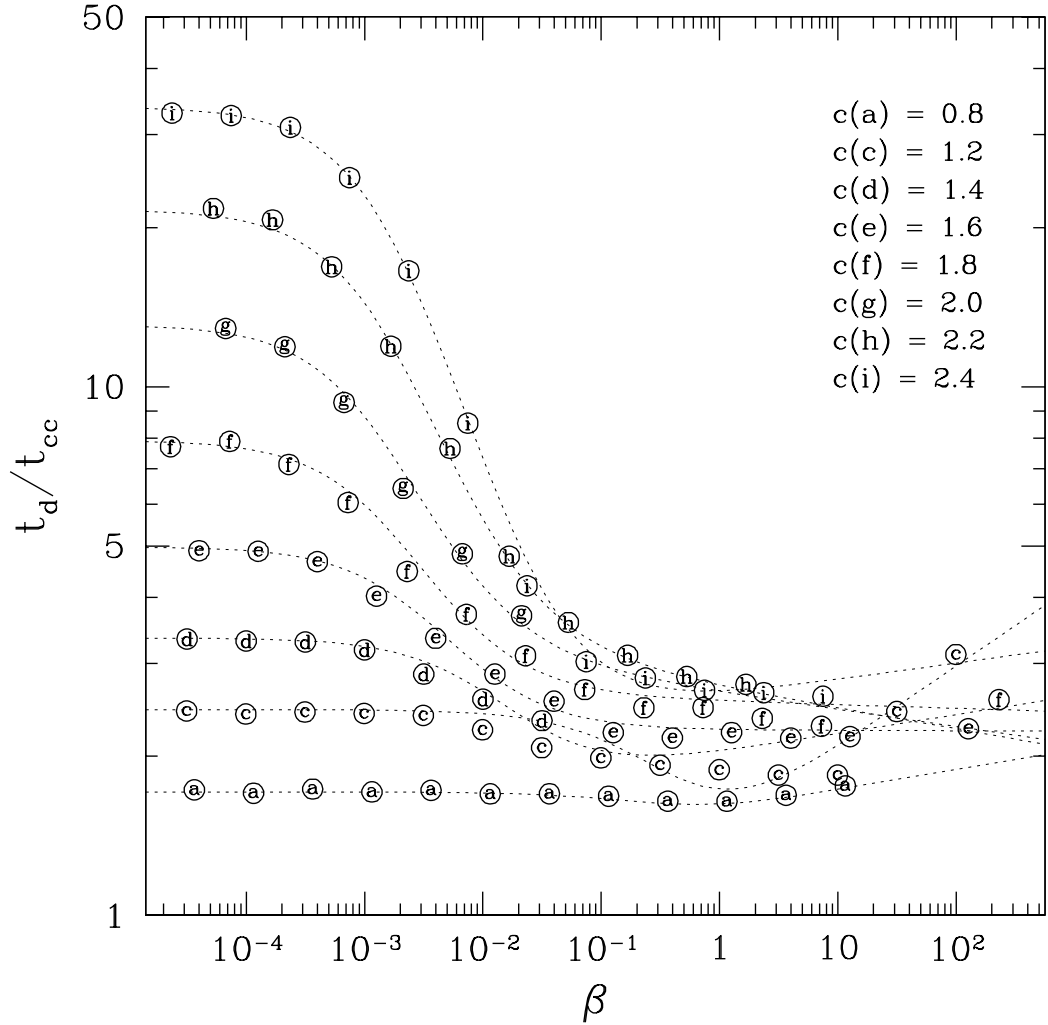}
\figcaption{The destruction time as a function of the shock parameter,
  $\beta$.  For clarity, some of the models are shown only for $\beta
  \lesssim 0.03$.  Dotted lines are our fit,
  eq. (\protect\ref{eq:td_sh}).  \label{fig:td_sh}}
\end{figure}

For the models including tidal shocks (Figure \ref{fig:td_sh}), the
scaled destruction time varies very little for the low concentration
clusters and increasingly more for the high concentration ones.
Interestingly enough, regardless of the initial structure, models with
strong shocks converge to an almost constant value $t_d/t_{cc} \sim 2$.
It follows that the core collapse time is a midpoint in cluster's
lifetime.

Since the value of $t_d/t_{cc}$ is almost constant for $\beta \ll 1$ and
for $\beta \gg 1$, we choose the following functional form to fit the
results:
\begin{equation}
{t_d \over t_{cc}} = \left. {t_d \over t_{cc}}\right|_{\beta=0} \;
  \left( {1 + d_1 \beta^{d_2} \over 1 + d_3 \beta} \right),
  \label{eq:td_sh}
\end{equation}
where the coefficients $a_i$ are given in Table \ref{tab:fittdbeta}.
This expression allows for a variation in the slope of the fit for large
values of $\beta$, but the coefficient $d_2$ is always close to unity,
as expected.

In the case of very strong shocks, both the core collapse and the
destruction times scale approximately as $t_{cc},t_d \propto
\beta^{-1/2}$ (cf. the coefficient $b_2$ in Table \ref{tab:fittccbeta}).
This asymptotic behavior can be understood as follows.  The shock
parameter $\beta$ is proportional to the mean energy change due to
shocks (eq. [\ref{eq:definetimes}]), which in turn is proportional to
the amplitude of the tidal force squared, $I^2$.  Therefore, the
approximate scaling of $t_{cc}$ and $t_d$ indicate that these timescales
vary inversely proportional to $I$.



\section{Discussion}
\label{sec:discussion}

Our Fokker-Planck models provide the most comprehensive calculations of
the globular cluster evolution including gravitational tidal shocks.  We
find that tidal shocks significantly alter the evolution, in agreement
with the earlier studies by \cite{SC:73} and more recent ones by
\cite{GO:97} (1997) and \cite{MW:97} (1997a-c).  Globular clusters
appear to be rather fragile with regard to the effects of tidal
shocking.

Tidal shocks accelerate dynamical evolution and evaporation of clusters
by effectively stripping stars in the outer regions and adding energy
dispersion in the core.  These effects are self-limiting: as clusters
lose mass and become more compact, the relative importance of tidal
shocks rapidly diminishes in favor of two-body relaxation.  In the case
of the Galactic globular cluster \ngc, tidal shocks decrease the
destruction time from 24 Gyr to 18 Gyr.

We examine several types of adiabatic corrections: the harmonic
approximation (of L. Spitzer), the linear perturbation theory (of
M. Weinberg), and the results of the \N-body simulations.  The
differences between the various types of adiabatic corrections depend on
the regime of ``slow shocks'' or ``fast shocks'', but overall the
corrections are critical for accurate calculation of the evolution.
Without adiabatic corrections, the destruction time of \ngc\ is twice as
short.

The evolution of models with a single mass component can be
characterized by the King-model concentration parameter, $c$, and the
shock parameter, $\beta$ (eq. [\ref{eq:betadef}]).  The effects of tidal
shocks are important when $\beta \gtrsim 10^{-3} - 10^{-2}$, depending
on the concentration.  Tidal shocks increase both the mean and the
central densities, relative to the relaxation case.  Core collapse is
accelerated by about 30\% for $\beta=0.01$ and by a factor of 3 for
$\beta=1$.  In the limit of strong shocks, the core collapse time scales
approximately as $t_{cc} \propto \beta^{-1/2}$.  The time to destruction
is also dramatically reduced, following the same scaling law.  Overall,
we find that evolution of models proceeds very similarly when time is
expressed in units of the core collapse time.  In the limit of strong
shocks, the destruction time $t_d \approx 2\, t_{cc}$, independently on
the initial concentration.  Core collapse provides a useful midpoint of
cluster evolution.

We also consider an illustrative model with a range of stellar masses.
In units of the initial relaxation time, the evolution proceeds much
faster than in the single-mass case, but for both models the time to
destruction is comparable when expressed in years.  Thus, the models
with the same average density evolve in approximately the same time.  A
systematic study of multi-mass models is beyond the scope of the present
study and requires a knowledge of the initial mass function in star
clusters.

The results of our calculations have important implications for the
evolution of globular clusters.  For the sample of the Galactic globular
clusters from \cite{GO:97} (1997) the typical values of the shock
parameter are $\beta \sim 10^{-3}-10^{-2}$.  This indicates that the
effects of tidal shocks and of two-body relaxation are comparable at
present.  Since the shock effects quickly saturate, tidal shocks must
have been more important in the past.  Not only will the current
population of clusters be significantly depleted in the future
(\cite{GO:97} 1997), but also a large fraction of the initial population
might have been already destroyed in the Galaxy (\cite{MW:97} 1997c).

Present calculations can be improved in several ways.  A reasonable mass
function and effects of stellar evolution should be included in order to
accurately describe the early phase of cluster evolution.  Also, in our
treatment of tidal shocks we have neglected small (but real) negative
relaxation due to cluster oscillations.  The final weakness of our
models is the assumption of the isotropic velocity distribution.  Radial
anisotropy would develop in the outer parts in the absence of tidal
shocks.  Since the stars on radial orbits will be ejected more easily
than those on circular orbits, tidal shocks should effectively
isotropize orbits even in the outer parts.  \cite{T} (1995,1997) and
\cite{Tetal:97} have developed an accurate method for integrating the
anisotropic Fokker-Planck equation.  We plan to include the effect of
tidal shocks in the anisotropic code in order to clarify these points.

\acknowledgements 

We would like to thank Kathryn Johnston for useful discussions, and the
anonymous referee and the scientific editor Steven Shore for detailed
comments.  This work was supported in part by NSF grant AST 94-24416 and
by Korea Science and Engineering Foundation grant No. 95-0702-01-01-3.


\begin{deluxetable}{lccccc}
\tablecaption{Model parameters for \ngc.
              \label{tab:param}}
\tablecolumns{6}
\tablewidth{0pt}
\tablehead{\colhead{Model} & \colhead{$c$} & \colhead{$t_{rh,0}$ (Gyr)} &
           \colhead{$t_{sh}/t_{rh,0}$} & \colhead{$t_{cc}/t_{rh,0}$} &
           \colhead{$t_d/t_{rh,0}$}}
\startdata
{\sc Single mass}       & 1.4\phn & 0.76 & 96 \nl
{\it case 0}            &         &      &     &     12.9 \phn & 32 \phn \nl
{\it case 1}            &         &      &     &     10.0 \phn & 26 \phn \nl
{\it case 2}            &         &      &     &     11.2 \phn & 32 \phn \nl
{\it case 3}            &         &      &     & \phn 9.7 \phn & 24 \phn \nl
 \tableline
{\sc Low concentration} & 0.84    & 1.80 & 7.3 \nl
{\it case 0}            &         &      &     &     12.2 \phn & 16.5 \nl
{\it case 1}            &         &      &     & \phn 8.0 \phn & 13.5 \nl
{\it case 2}            &         &      &     & \phn 8.5 \phn & 14.5 \nl
{\it case 3}            &         &      &     & \phn 7.1 \phn & 12 \phn \nl
 \tableline
{\sc Multi-mass}        & 1.4\phn & 2.77 & 15 \nl
{\it case 0}            &         &      &     & \phn 1.76     & 13.5 \nl
{\it case 1}            &         &      &     & \phn 1.59     & \phn 9.5 \nl
{\it case 2}            &         &      &     & \phn 1.60     & 13 \phn \nl
{\it case 3}            &         &      &     & \phn 1.57     & \phn 8.5
\enddata
\end{deluxetable}

\begin{deluxetable}{lccccc}
\tablecaption{Fitting coefficients for $t_{cc}$ and $t_d$ versus $c$.
              \label{tab:fitc}}
\tablecolumns{6}
\tablewidth{0pt}
\tablehead{\colhead{Function} & \colhead{$a_1$} & \colhead{$a_2$} &
           \colhead{$a_3$} & \colhead{$a_4$} & \colhead{$a_5$}}
\startdata
$f_1(c)$ & 1.3162 & -1.9809 &  3.2910 & -1.8401 &  0.2987 \nl
$f_2(c)$ & 1.2897 & -1.1223 &  1.9824 & -0.8803 &  0.1104 \nl
$f_3(c)$ & 0.1286 &  0.3173 & -0.6573 &  0.6367 & -0.1319
\enddata
\end{deluxetable}

\begin{deluxetable}{lcccc}
\tablecaption{Fitting coefficients for $t_{cc}$ versus $\beta$.
              \label{tab:fittccbeta}}
\tablecolumns{5}
\tablewidth{0pt}
\tablehead{\colhead{$c$} & \colhead{$b_1$} & \colhead{$b_2$} &
           \colhead{$b_3$} & \colhead{$b_4$}}
\startdata
0.8 & 1.5029 & 0.5539 &  0.0185 & 0.4558 \nl
1.0 & 0.1174 & 1.4184 &  1.6004 & 0.4491 \nl
1.2 & 0.5124 & 0.8597 &  1.5189 & 0.3840 \nl
1.4 & 3.3268 & 0.4680 & -1.0819 & 0.5281 \nl
1.6 & 2.9924 & 0.4613 & -1.0366 & 0.5726 \nl
1.8 & 1.1447 & 0.3691 &  0.1791 & 0.3614 \nl
2.0 & 1.4912 & 0.2996 & -0.7153 & 0.1606 \nl
2.2 & 0.8931 & 0.6518 & -0.6721 & 0.2626 \nl
2.4 & 0.6413 & 0.5716 & -0.9622 & 0.2167 \nl
2.6 & 0.9452 & 0.5916 & -1.4586 & 0.2065
\enddata
\end{deluxetable}

\begin{deluxetable}{lccc}
\tablecaption{Fitting coefficients for $t_d$ versus $\beta$.
              \label{tab:fittdbeta}}
\tablecolumns{4}
\tablewidth{0pt}
\tablehead{\colhead{$c$} & \colhead{$d_1$} & \colhead{$d_2$} &
           \colhead{$d_3$}}
\startdata
0.8 & 2.1942 & 1.0376 & 2.3645 \nl
1.0 & 0.1315 & 1.2982 & 0.2646 \nl
1.2 & 1.6051 & 1.1528 & 2.6816 \nl
1.4 & 45.420 & 1.0356 & 74.731 \nl
1.6 & 135.52 & 0.9993 & 300.84 \nl
1.8 & 188.62 & 0.9932 & 587.29 \nl
2.0 & 159.24 & 0.9675 & 788.60 \nl
2.2 & 83.444 & 0.9608 & 667.98 \nl
2.4 & 40.040 & 1.0309 & 520.84 \nl
2.6 & 42.112 & 1.1056 & 862.75 \nl
\enddata
\end{deluxetable}

\clearpage


\clearpage

\end{document}